\documentclass{tacmconf}
\usepackage{graphicx} 
\usepackage{epsfig}
\usepackage{stmaryrd}
\usepackage{latexsym}
\usepackage{amssymb}
\usepackage{daytime}
\usepackage{eepic}
\usepackage{url}
\usepackage{verbatim}
\usepackage{wasysym}
\def\thisPaperTitle{On Role Logic}

\usepackage{defs}
\title{\thisPaperTitle}

\author{Viktor Kuncak and Martin Rinard \\
        Computer Science and Artificial Intelligence Laboratory\\
        Massachusetts Institute of Technology\\
        Cambridge, MA 02139, USA \\
        {\tt $\{$vkuncak,rinard$\}$@csail.mit.edu} \\ \\
        {MIT CSAIL Technical Report No 925} \\
        {Internal Manuscript VK0101, October 2003}
}

\begin{document}

\sloppy

\maketitle

\renewcommand{\thefootnote}{\fnsymbol{footnote}}
\footnotetext{Draft of \today, \Daytime, \\ see 
\url{http://www.mit.edu/~vkuncak/papers} for later versions.}

\renewcommand{\thefootnote}{\arabic{footnote}}

\begin{abstract}
We present {\em role logic}, a notation for describing
properties of relational structures in shape analysis,
databases, and knowledge bases.  We construct role logic
using the ideas of de Bruijn's notation for lambda calculus,
an encoding of first-order logic in lambda calculus, and a
simple rule for implicit arguments of unary and binary
predicates.

The unrestricted version of role logic has the expressive
power of first-order logic with transitive closure.  Using a
syntactic restriction on role logic formulas, we identify a
natural fragment $\RLtwo$ of role logic.  We show that the
$\RLtwo$ fragment has the same expressive power as
two-variable logic with counting $C^2$, and is therefore
decidable.

We present a translation of an imperative language into the
decidable fragment $\RLtwo$, which allows compositional
verification of programs that manipulate relational
structures.  In addition, we show how $\RLtwo$ encodes
boolean shape analysis constraints and an expressive
description logic.
\end{abstract}




\pagebreak
\tableofcontents
\pagebreak

\section{Introduction}


\smartparagraph{Systems as relational structures.} Complex
systems arising in many areas of Computer Science can be
naturally represented as relational structures.  The state
of an imperative program can be specified using sets and
relations denoted by unary and binary predicates
\cite{Floyd67AssigningMeaningsPrograms,
  Hoare69AxiomaticBasisComputerProgramming,
  SchonbergETAL91Setl,
  BoergerStaerk03AbstractStateMachines}, especially for
object-oriented programs \cite{Jackson02AlloyTOSEM,
  RumbaughETAL99UMLReference}; a relational database is a
finite relational structure
\cite{Codd70RelationalModelDataLargeSharedDataBanks,
  Chen76EntityRelationshipModel}; knowledge bases and
deductive databases can also be based on predicate logic
\cite{BaaderETAL03DescriptionLogicHandbook,
  Kowalski79AlgorithmLogicControl,
  Lloyd87FoundationsLogicProgramming}.

\smartparagraph{Shape analysis.} Shape analysis techniques
\cite{SagivETAL02Parametric, GhiyaHendren96TreeOrDag,
  HummelETAL94GeneralDataDependence,
  FradetMetayer97ShapeTypes, 
  FradetMetayer98StructuredGamma,
  FradetETALStaticVerificationPointer,
  ChongRugina03StaticAnalysisAccessedRegionsRecursiveDataStructures,
  KlarlundSchwartzbach94Transductions,
  KlarlundSchwartzbach93GraphTypes,
  KuncakETAL02RoleAnalysis,
  JensenETAL97MonadicLogic, Moeller01PALE} can verify and
derive precise properties of objects in the heap.  Shape
analysis is therefore important for reasoning about programs
written in modern imperative programming languages.  Shape
analysis is also promising as a general-purpose verification
technique, because of its ability to reason about graphs as
general structures, and the ability to summarize properties
of unbounded sets of objects.

Many of the shape analysis techniques have a logical
foundation: \cite{SagivETAL02Parametric} is based on
(two-valued and three-valued) first-order logic with
transitive closure, \cite{KlarlundSchwartzbach93GraphTypes,
  KlarlundSchwartzbach94Transductions,
  JensenETAL97MonadicLogic, Moeller01PALE} is based on
monadic second-order logic of trees,
\cite{FradetMetayer97ShapeTypes,
  FradetMetayer98StructuredGamma} is based on graph grammars
which are closely related to monadic second-order logic of
trees \cite{Rozenberg97HandbookGraphGrammars}.  Theorem
proving is used in \cite{HummelETAL94GeneralDataDependence}
to derive consequences of axioms about data structures.
Many shape analyses perform abstract interpretation
\cite{CousotCousot77AbstractInterpretation} to synthesize
loop invariants \cite{SagivETAL02Parametric,
  GhiyaHendren96TreeOrDag, KuncakETAL02RoleAnalysis}.

\smartparagraph{Role logic.}  This paper presents {\em role
  logic}, a notation for describing properties of relational
structures in shape analysis, databases, and knowledge bases.
Role logic is an attempt to simultaneously achieve the
simplicity of the role declarations of
\cite{KuncakETAL02RoleAnalysis} with a transparent
connection with the well-established first-order logic.

On the one hand, the full role logic has the expressive
power of first order logic with transitive closure, which
makes it as expressive as the logic of
\cite{SagivETAL02Parametric, Jackson02AlloyTOSEM} and more
expressive than the original role constraints
\cite{KuncakETAL02RoleAnalysis}.  For example, role logic is
closed under all propositional operations and generalizes
boolean shape analysis constraints
\cite{KuncakRinard03OnBooleanAlgebraShapeAnalysisConstraints}.
Role logic formulas easily translate into the traditional
first-order logic notation.

On the other hand, like the specialized notation for
declaring roles in \cite{KuncakETAL02RoleAnalysis}, role
logic allows natural description of common properties of
imperative data structures with mutable references.  Like
dynamic logics \cite{HarelETAL00DynamicLogic} and
description logics
\cite{BaaderETAL03DescriptionLogicHandbook}, role logic
allows suppressing names of variables, which often leads to
concise specifications.  The conciseness of role logic makes
it an appealing choice for lightweight annotations in a
programming language.

Another property that role logic shares with description
logics is that an interesting subset of role logic is
decidable.  We show the decidability of the fragment
$\RLtwo$ of role logic in Section~\ref{sec:decidable} by
establishing a correspondence with the two-variable logic
with counting $C^2$
\cite{GraedelETAL97TwoVariableLogicCountingDecidable,
  PacholskiETAL00ComplexityResultsFirstOrderTwo}.  While
many description logics are known to be representable in
$C^2$ but are potentially weaker than $C^2$, the fragment
$\RLtwo$ of role logic matches precisely the expressive
power of $C^2$.

\smartparagraph{Contributions.}
The following are the main contributions of this paper:
\begin{enumerate}
\item We introduce \emph{role logic}, which applies the
  ideas of implicit arguments and deBruijn's lambda calculus
  notation to first order logic (Section~\ref{sec:recipe}).
  The result is a concise way of specifying properties of
  first-order structures that arise in shape analysis,
  databases, and knowledge bases.
\item We define a variable-free subset $\RLtwo$ of role
  logic (Section~\ref{sec:decidable}).  We give a
  translation of $\RLtwo$ formulas to formulas of
  two-variable logic with counting $C^2$.  This translation
  implies that $\RLtwo$ is decidable, because $C^2$ is
  decidable
  \cite{GraedelETAL97TwoVariableLogicCountingDecidable}.  We
  further give a translation of $C^2$ formulas to $\RLtwo$
  formulas.  These two translations imply that $\RLtwo$ is
  just as expressive as $C^2$.
\item As the main application of role logic, in
  Section~\ref{sec:transductions} we present a
  compositional shape analysis technique.  We introduce a
  unified language for writing implementations,
  specifications, and conformance claims.  The constructs of
  the language denote relations on program states
  expressible in the decidable fragment $\RLtwo$.  The
  analysis technique is based on generating verification
  conditions in $\RLtwo$ and applying the decision procedure
  for $\RLtwo$.  The analysis verifies the correctness of
  the dynamically changing referencing relationships between
  objects by showing that procedures conform to their
  specifications.  By conjoining procedure specifications
  with global invariants, the analysis can also show that
  the program preserves the key data structure consistency
  properties necessary for the correct execution of the
  program.
\item We present two additional applications of role logic:
  \begin{enumerate}
  \item we show in Section~\ref{sec:dlencoding}
    that a subset of role logic $\RLtwo$ naturally
    corresponds to an expressive description logic
    \cite[Chapter 5]{BaaderETAL03DescriptionLogicHandbook};
  \item we note in Section~\ref{sec:bsac} that boolean shape
    analysis constraints
    \cite{KuncakRinard03OnBooleanAlgebraShapeAnalysisConstraints},
    which can describe the basic structure of data-flow
    facts in \cite{SagivETAL02Parametric}, are a subset of
    constraints expressible in role logic.
  \end{enumerate}
\end{enumerate}



\section{Example}
\label{sec:example}

\begin{figure}[ht]
\begin{center}
\includegraphics[height=1.2in]{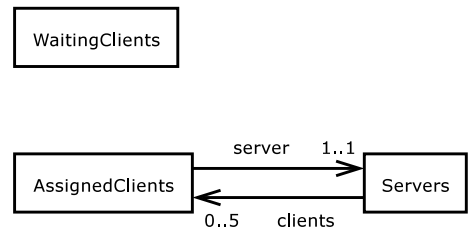}
\end{center}
\caption{An object model for a component of client-server manager\label{fig:exampleObjectModel}}
\end{figure}

To give a flavor of role logic, we present an example that
illustrates one aspect of a client-server manager system
that assigns clients to servers.
Figure~\ref{fig:exampleObjectModel} is a standard object
model that graphically displays the system, using boxes to
represent sets, arrows to represent relations, and intervals
$N..M$ to represent constraints on relations.
Figure~\ref{fig:exampleText} describes the same system using
role logic.
Figure~\ref{fig:exaProgram} presents a fragment of the code
of the system.  The code is expressed in an imperative
language extended with specification constructs.




\begin{figure}[ht]
  \begin{equation*}
    \begin{array}{l}
      \GlobalInvariant = {} \mnl
      \begin{array}{@{\quad}l}
        \curlyb{\Servers} \ \land \
        (\disjoint\, \Servers, \Clients) \ \land \mnl
        (\partition\, \Clients; \ \WaitingClients, \AssignedClients)\ \land \mnl
        \squareb{\squareb{\server \implies \AssignedClients' \land \Servers}}\ \land \mnl
        \squareb{\squareb{\clients \miff \twid{\server}}}\ \land \mnl
        \squareb{\AssignedClients \implies \card{{=}1} \server}\ \land \mnl
        \squareb{\Servers \implies \card{{\leq}5} \clients}
      \end{array} \\
      \\
      \mbox{Example consequence:} \\
      \\
      P \ \equiv \
      \squareb{\WaitingClients \implies \mnl
        \qquad\qquad
          \squareb{\lnot (\clients \lor \server \lor 
                  \twid{\clients} \lor \twid{\server})}}
    \end{array}
  \end{equation*}
  \caption{Global constraints of the client-server manager, expressed in role logic
    \label{fig:exampleText}}
\end{figure}

\smartparagraph{Global constraints.}
Figure~\ref{fig:exampleText} describes the global
constraints of a client-server manager system using a
conjunction of role logic formulas.  There are two basic
kinds of objects in the system: servers and clients.  We
model these objects using two disjoint sets $\Clients$ and
$\Servers$.  The set $\Clients$ is further partitioned into
the set $\AssignedClients$ of objects that have been
assigned to servers, and the set $\WaitingClients$ that have
not been assigned yet.  The $\disjoint$, $\partition$, and
other constructs of set algebra of sets and relations
($\cap$, $\cup$, $\setminus$) are definable in role logic.

We require the set $\Servers$ to be non-empty, which we
denote by $\curlyb{\Servers}$, with the meaning $\exists x.
\Servers(x)$.  The constraint $\squareb{\squareb{\server
    \implies \AssignedClients' \land \Servers}}$ translates
to $\forall x. \forall y.\, \server(x,y) \implies
\AssignedClients(x) \land \Servers(y)$.  Namely, the
brackets $\squareb{\ }$ corresponds to a universal
quantifier.  An occurrence of a binary predicate (such as
$\server$) is implicitly supplied with the
previous-innermost bound variable (here, $x$) and the innermost
bound variable (here, $y$).  The occurrence of an unary predicate
$\Servers$ is supplied with the innermost bound variable
($y$), unless the unary predicate is primed, in which case the
previous-innermost bound variable (in this case $x$) is
supplied instead.  The constraint
$\squareb{\squareb{\clients \miff \twid{\server}}}$ means
that the relation $\clients$ is the inverse of the relation
$\server$.  The constraint $\squareb{\Servers \implies
  \card{{\leq}5} \clients}$ translates into the formula
$\forall x.\, \Servers(x) \implies
\existsleq{5}{y}{\clients(x,y)}$ in first-order logic
with counting quantifiers.

Note that all of our translations of constraints in
Figure~\ref{fig:exampleText} use only two variables, $x$ and
$y$.  In fact, our entire example is expressed in the
$\RLtwo$ fragment of role logic.  In
Section~\ref{sec:decidable} we show that $\RLtwo$
corresponds to the decidable fragment $C^2$ of two-variable
first-order logic with counting, and is therefore decidable.
Figure~\ref{fig:exampleText} presents the formula $P$
denoting the fact that $\WaitingClients$ objects have no
incoming or outgoing edges.  If we apply the decision
procedure for $\RLtwo$, we can show that $\GlobalInvariant
\implies P$ is a valid formula, which means that $P$ is a
logical consequence of $\GlobalInvariant$.  By querying
whether the $\GlobalInvariant$ implies properties of
interest such as $P$, the developers can increase their
confidence in the correctness and completeness of the
design.  Moreover, our technique can be used to show the
conformance of the program with respect to the design.

\begin{figure}
\begin{verbatim}
proc assignClients() = 
spec old(GlobalInvariant) => !{WaitingClients} & 
 [AssignedClients <=> 
        old(AssignedClients | WaitingClients)] &
 GlobalInvariant

proc assignClientsIMPL() = {
  if ({WaitingClients}) {
    cl := getWaitingClient();
    assignOneClientIMPL(cl);
    assignClientsIMPL();
}}
claim: assignClientsIMPL => assignClients

proc assignOneClient(cl) =
spec old(GlobalInvariant & 
     [cl => WaitingClients]) =>
 [WaitingClients | cl <=> old(WaitingClients)] &
 [AssignedClients <=> old(AssignedClients) | cl] &
 GlobalInvariant

proc assignOneClientIMPL(cl) = {
  sv := getServer();
  if (Card (sv' & clients) <= 4) {
    WaitingClients := WaitingClients \ cl;
    AssignedClients := AssignedClients | cl;
    cl.server := sv;
    sv.clients := sv.clients | cl;
  } else {
    assignOneClientIMPL(cl);
}}
claim: assignOneClientIMPL => assignOneClient

proc getWaitingClient() : set =
  spec {WaitingClients} =>
       skip & [returned => WaitingClients]

proc getServer() : set =
  spec {Servers} =>
       skip & [returned => Servers]
\end{verbatim}
\caption{A fragment of a program that assigns $\WaitingClients$ to $\Servers$\label{fig:exaProgram}}
\end{figure}

\smartparagraph{Program fragment.}
Figure~\ref{fig:exaProgram} shows a fragment of the code of
the client-server manager.  The top-level procedure in the
code is a tail-recursive procedure $\assignClientsIMPL$ that
processes all $\WaitingClients$ objects and assigns them to
$\Servers$ objects.  The $\assignClientsIMPL$ procedure
terminates if there are no $\WaitingClients$ objects.
Otherwise, it uses the $\getWaitingClient$ procedure to
obtain an element of $\WaitingClients$ and assigns it to
some $\Servers$ object using the $\assignOneClient$
procedure, and continues with the next $\WaitingClients$
object using a tail-recursive call.

The partial correctness of the procedure
$\assignClientsIMPL$ is given using the specification
$\assignClients$.  The requirement that the procedure
conforms to its specification is stated using the construct
\begin{verbatim}
claim: assignClientsIMPL => assignClients
\end{verbatim}
The verification of each procedure call site uses only
procedure specification (summary) instead of the body of the
procedure, which allows verification of recursive
procedures.  In this example, the implementations of
procedures $\getWaitingClient$ and $\getServer$ are
not available, which illustrates the advantage of
assume/guarantee reasoning for partitioning a verification
task.

Using the translation in Section~\ref{sec:transductions},
the $\q{claim}$ constructs are reduced to verification
conditions expressed in role logic.  For a large class of
constructs presented in Section~\ref{sec:transductions}, and
our example in particular, the resulting verification
conditions belong to the decidable $\RLtwo$ and can
therefore be discharged using a decision procedure for
$\RLtwo$.

Note that we are able to express detailed specifications of
the correctness of procedures while remaining in the
decidable logic.  For example, the specification
$\assignClients$ ensures that the entire global invariant in
Figure~\ref{fig:exampleText} is preserved, and that no
client objects are lost in the assignment process: after
$\assignClients$, the set $\AssignedClients$ is the union of
the old value of $\AssignedClients$ and the old value of
$\WaitingClients$, whereas the new value of
$\WaitingClients$ is an empty set.



\section{A Recipe for Role Logic}
\label{sec:recipe}

In this section we motivate the role logic by constructing
it in several steps.  We start with first-order logic
encoded in the simply typed lambda calculus; we then move to
the notation that refers to each variable by its index.
Finally, we impose a rule for implicitly supplying the
indices of variables to predicate symbols.  Later, in
Section~\ref{sec:roleLogic}, we summarize the syntax and the
semantics of role logic, and in Section~\ref{sec:decidable}
we present a decidable sublogic of role logic.

\subsection{Lambda Calculus}

\begin{figure}
  \begin{equation*}
    \begin{array}{rcll}
      \Formula &=& \Vars
      & \begin{array}[t]{@{}l}
        \mbox{variable lookup} \\ 
        \Vars = \{ x, f, \ldots \} \mnl
      \end{array} \\
      & \mid & \Formula\ \Formula 
      & \cmnt{function application} \mnl
      & \mid & \lambda \Vars : \Type . \Formula
      & \cmnt{function abstraction} \mnl
    \end{array}
  \end{equation*}
  \centerline{\bf Syntax}
  \centerline{}
  \begin{equation*}
    \begin{array}{c}
      \trule{\Gamma(v)=T}
      {\Gamma \vdash v:T} \mnl
      \\
      \trule{\Gamma \vdash F_1 : T_1 \to T_2, \quad \Gamma \vdash F_2 : T_1}
      {\Gamma \vdash F_1 F_2 : T_2} \mnl
      \\
      \trule{\Gamma[v:=T_1] \vdash F:T_2}
      {\Gamma \vdash (\lambda v : T_1 . F) : T_1 \to T_2} \mnl
    \end{array}
  \end{equation*}
  \centerline{\bf Types}
  \centerline{}
  \begin{equation*}
    \begin{array}{rcl}
      \tr{v}\, e &=& e\, v \mnl
      \tr{F_1\ F_2}\, e &=& (\tr{F_1}e)\ (\tr{F_2}e) \mnl
      \tr{\lambda v : T . F}\, e 
      &=& \lambda d . \tr{F}\, (e[v:=d])
    \end{array}
  \end{equation*}
  \centerline{\bf Semantics}
  \caption{Church-style Simply Typed Lambda Calculus
    \label{fig:simpleChurch}}
\end{figure}

Figure~\ref{fig:simpleChurch} presents simply typed lambda
calculus with explicit type annotations in lambda
abstraction (the Church-style simply typed lambda calculus
\cite[Section 3.2]{Barendregt01LambdaCalculiTypes}).  This
calculus is our starting point.  

As primitive types we use
$\bool$ for boolean values, and $\obj$ for objects.  As the
only type constructor we use arrow $\to$.  We introduce $\reltp{k}$
as a shorthand type defined by
\begin{equation*}
  \begin{array}{rcl}
    \reltp{0} & \equiv & \bool \mnl
    \reltp{k+1} & \equiv & \obj \to \reltp{k}
  \end{array}
\end{equation*}
Simple types enable us to give a simple set-theoretic
semantics to formulas by interpreting lambda abstractions as
total functions.  The resulting semantics is in
Figure~\ref{fig:simpleChurch}; the semantics is
straightforward because we use lambda calculus itself as our
meta-notation.

\subsection{De Bruijn Notation}

\begin{figure}
  \begin{equation*}
    \begin{array}{rcll}
      \Formula &=& \ind{\Nat}
      & \begin{array}[@{}t]{l}
        \mbox{variable lookup} \\
        \Nat = \{ 1,2,\ldots\} \mnl
      \end{array} \\
      & \mid & \Formula\ \Formula 
      & \cmnt{function application} \mnl
      & \mid & \lambda\, {:}\Type . \Formula
      & \cmnt{function abstraction} \mnl
    \end{array}
  \end{equation*}
  \centerline{\bf Syntax}
  \centerline{}
  \begin{equation*}
    \begin{array}{rcl}
      \tr{\ind{i}}\, e &=& \sget\, i\, e \mnl
      \tr{F_1\ F_2}\, e &=& (\tr{F_1}e)\ (\tr{F_2}e) \mnl
      \tr{\lambda\, {:} T . F}\, e 
      &=& \lambda d .\ \tr{F}\, (\push\, d\, e)
    \end{array}
  \end{equation*}
  \centerline{\bf Semantics}
  \centerline{}
  \begin{equation*}
    \begin{array}{rcl}
      \sget\, i\, e &=& \nth\, i\ (e\ \stack) \mnl
      \push\, d\, e &=& e[\stack := d : (e\, \stack)] \mnl
      \nth\, 1\, (h:l) &=& h \mnls
      \nth\, (i+1)\, (h:l) &=& \nth\, i\, l
    \end{array}
  \end{equation*}
  \centerline{\bf Auxiliary Functions}
  \caption{De Bruijn Form of Simply Typed Lambda Calculus
    \label{fig:deBruijn}}
\end{figure}

An alternative to referring to each bound variable by its
name is to refer to each variable by its number, with number
$1$ denoting the most recently bound variable.  This is the
idea behind de Bruijn indices for lambda calculus
\cite{Bruijn72LambdaCalculusNotationNamelessDummies,
  Barendregt84LambdaCalculus}.  Figure~\ref{fig:deBruijn}
presents the syntax and the semantics of lambda calculus
notation with de Bruijn indices.  The environment maps the
keyword {\em stack} to a stack (i.e., a list) of elements of
the domain.  If $h$ is an element and $l$ a list, then the
notation $h:l$ denotes the list with the head $h$ and the
tail $l$.  The abstraction pushes a value onto the stack;
the index $\ind{k}$ retrieves the $k$-th element from the
top of the stack.

\subsection{Predicate Logic in Lambda Calculus}

\begin{figure}
  \begin{equation*}
    \begin{array}{rcl}
      \ID     &::& \reltp{2} \mnl
      \tr{\ID}\, x\, y &=& (x=y) \mnl
      \land   &::& \bool \to \bool \to \bool \mnl
      \tr{\land}\, p\, q &=& p \land q \mnl
      \lnot   &::& \bool \to \bool \mnl
      \tr{\lnot}\, p &=& \lnot p \mnl
      \exists &::& \reltp{1} \to \bool \mnl
      \tr{\exists}\, f &=& \exists o \in \tr{\obj}.\ f\, o \mnl
      \\
      \exists v. F & \equiv & \exists (\lambda v:\obj.\ F) \mnl
      \forall v. F & \equiv & \lnot \exists v. \lnot F
    \end{array}
  \end{equation*}
  \caption{First-Order Logic in Lambda Calculus
\label{fig:holEncoding}}
\end{figure}

We next encode first-order logic with equality in lambda
calculus.  We use $\ID$ to denote the binary equality
relation.  We assume that the interpretation of relation
symbols is specified in the environment $e$.  We introduce
conjunction and negation as logical operations acting on
booleans (the remaining propositional operations are defined
in terms of $\land,\lnot$, as usual).  We use the
abstraction in lambda calculus to encode bound variables of
predicate calculus.  This is the usual higher-order logic
encoding of classical first-order logic, as used, for
example, in Isabelle interactive theorem prover
\cite{Paulson94Isabelle}.  Figure~\ref{fig:holEncoding}
presents this encoding of quantifiers.  To remain within
first-order logic, we require the quantifier $\exists$ to
have monomorphic type $(\obj \to \bool) \to \bool$ (see also
Section~\ref{sec:lambdaDefinitions}).

\subsection{Implicit De Bruijn Indices}

\begin{figure}
  \begin{equation*}
    \begin{array}{rcl}
      \curlyb{F}  & \equiv & \forall (\lambda\, {:}\obj . F) \mnl
      \squareb{F} & \equiv & \lnot \curlyb{\lnot F} \mnl      
    \end{array}
  \end{equation*}
  \centerline{\bf Quantifier Brackets}
  \centerline{}
  \begin{center}
    \begin{tabular}{rcl}
      When $\Gamma(r) = \reltp{k}$ & 
      then write &
      $r$ \\
      & instead of &
      $r \ind{k} \ind{k{-}1} \ldots \ind{1}$
    \end{tabular}
  \end{center}
  \centerline{\bf Default Argument Rule}
  \centerline{}
  \begin{equation*}
    \begin{array}{rcl}
      \twid{F} & \equiv & (\lambda \lambda F) \ind{1}\ind{2} \mnl
      F' & \equiv & (\lambda \lambda F)\ind{2}\ind{2} \mnl
      \card{{\geq}k} F & \equiv & 
      \{^k \begin{array}[t]{l}
             (\lambda F)\ind{1} \land \ldots \land (\lambda F)\ind{k} \land \mnl
             \bigwedge_{1 \leq i < j \leq k} \lnot \ID \ind{i}\ind{j}\ \}^k \mnl
           \end{array} \mnl
      \card{{=}k} F & \equiv &
        \card{{\geq}k} F \land \lnot \card{{\geq}k{+}1} F \mnl
      (\sum_{i=1}^n \fcard\, F_i)  \geq k &\equiv& 
      \bigvee\limits_{\sum_{i=1}^n k_i = k} \bigwedge_{i=1}^n \card{{\geq}k_i} F_i \mnl      
      (\sum_{i=1}^n \fcard\, F_i)  = k &\equiv& 
      \bigvee\limits_{\sum_{i=1}^n k_i = k} \bigwedge_{i=1}^n \card{{=}k_i} F_i \\
      \\      
      \disjoint\, F_1,\ldots,F_n & \equiv &
        [\bigwedge\limits_{1 {\leq} i {<} j {\leq} n}
          \lnot (F_i \land F_j)] \\
      \\
      \partition\, F;F_1,\ldots,F_n & \equiv &
        \begin{array}[t]{l}
           \disjoint\, F_1,\ldots,F_n\ \land \mnl
          [F \miff \bigvee_{i=1}^n F_i]
        \end{array} \\
     \\
     F_1 \setminus F_2 & \equiv & F_1 \land \lnot F_2
     \end{array}
  \end{equation*}
  \centerline{\bf Shorthands}
  \caption{de Bruijn form of Predicate Calculus
\label{fig:dbpc}}
\end{figure}

Figure~\ref{fig:dbpc} shows how we combine the encoding of
first-order logic in higher-order logic and de Bruijn's
notation for lambda calculus. 
\begin{example} \label{exa:dbpc}
  First-order predicate calculus formula
  \begin{equation*}
    \forall x \forall y.\ f(x,y) \Rightarrow A(x) \land B(y)
  \end{equation*}
  can be written in this notation as
  \begin{equation*}
    [[ f \ind{2} \ind{1} \implies A \ind{2}  \land B \ind{1} ]]
  \end{equation*}
  The outermost $[\ ]$ bracket acts as the quantifier
  $\forall x$; the variable $x$ is referred to inside the
  formula as $\ind{2}$ because it is the second innermost
  bound variable.  The innermost $[\ ]$ bracket acts as
  $\forall y$; the variable $y$ is referred to as
  $\ind{1}$.
\end{example}
The interpretation environment $e$ contains both the stack
for de Bruijn indices and the bindings of relation symbols
such as $A$ and $f$ in Example~\ref{exa:dbpc}.  Relation
symbols of predicate logic correspond to variables of type
$\reltp{k}$.  We use the abstraction over de Bruijn indices
$\lambda\, {:}T. F$ only when $T \equiv \obj$, and write
this abstraction simply $\lambda F$.  For every environment
$e$, the value $(e\, \stack)$ is a list of elements of type
$\obj$.

We next introduce the Default Argument Rule: we omit de
Bruijn indices from the expression $r \ind{k} \ind{k{-}1}
\ldots \ind{1}$ when $r$ is a relation symbol, that is, when
$\Gamma(r)=\reltp{k}$.  We interpret every occurrence of
variable $r$ when $\Gamma(r)=\reltp{k}$ as $r \ind{k}
\ind{k{-}1} \ldots \ind{1}$.
\begin{example} \label{exa:dbpcDefault}
  The Default Argument Rule means that instead of
  \begin{equation*}
    [[ f \ind{2} \ind{1} \implies A \ind{2}  \land B \ind{1} ]]  
  \end{equation*}
  we write
  \begin{equation*}
    [[ f \implies (\lambda A)\ind{2} \land B ]]
  \end{equation*}
  when $\Gamma(f)=\reltp{2}$ and $\Gamma(A)=\Gamma(B)=\reltp{1}$.
\end{example}
We lose no expressive power by the Default Argument Rule.
For example, if we wish to denote $r \ind{i_3} \ind{i_2}
\ind{i_1}$, we write $(\lambda \lambda \lambda r) \ind{i_3}
\ind{i_2} \ind{i_1}$.  Note that the Default Argument Rule
applies only to the relation symbols, not to all
subformulas, so $(\lambda \lambda \lambda r)$ with Default
Argument rule is equivalent to $r$ without Default Argument
Rule.  In general, if $r$ is an $n$-ary relation, we write
$((\lambda)^k r)\ind{i_k} \ind{i_{k-1}} \ldots \ind{i_1}$
where we would previously write $r \ind{i_k} \ind{i_{k_1}}
\ldots \ind{i_1}$.

\subsection{Shorthands}

Figure~\ref{fig:dbpc} introduces some shorthands.  Tilde
$\twid{}$ swaps two topmost stack elements $\ind{1}$ and
$\ind{2}$. Prime $'$ replaces the top $\ind{1}$ with the
element $\ind{2}$.  An expression $\card{{\geq}k} F$, for an
integer $k \geq 0$, corresponds to a counting quantifier in
first-order logic
\cite{GraedelETAL97TwoVariableLogicCountingDecidable}.  A
counting quantifier states that the number of elements with
some property is greater than or equal to $k$.
Figure~\ref{fig:dbpc} also introduces the shorthand for
$\card{{=}k} F$ and the shorthand $\fcard$ for specifying a
constraint on a sum of cardinalities.  The shorthands
containing $\leq$ are defined similarly.

These shorthands play two purposes.  On the one hand they
allow expressing certain properties in a more concise way.
On the other hand, if we use the shorthands but give up the
ability to refer to indices explicitly, we obtain a fragment
of first-order logic that is equivalent to two-variable
first-order logic with counting
(Section~\ref{sec:decidable}) and therefore decidable
\cite{GraedelETAL97TwoVariableLogicCountingDecidable}.
\begin{example} \label{exa:withShorthands}
  Using the shorthands, we write the formula
  \begin{equation*}
    \forall x \forall y.\ f(x,y) \Rightarrow A(x) \land B(y)
  \end{equation*}
  as
  \begin{equation*}
    [[ f \implies A' \land B ]]
  \end{equation*}
  The convenience of role logic is even more evident in larger
  formulas like
  \begin{equation*}
    \forall x.\ A(x) \Rightarrow
    \begin{array}[t]{@{}l}
      (\forall y. f(x,y) \Rightarrow B(y) \lor C(y)) \ \land \mnl
      (\forall z. g(x,z) \Rightarrow D(z))
    \end{array}
  \end{equation*}
  which can be written as
  \begin{equation} \label{eqn:pointsToFormulaExample}
    [A \implies [f \implies B \lor C] \land [g \implies D]]
  \end{equation}
  Formulas of form~(\ref{eqn:pointsToFormulaExample}) are
  useful for describing properties of first order structures
  that arise in shape analysis, see e.g.\ 
  \cite{KuncakRinard03OnBooleanAlgebraShapeAnalysisConstraints,
    KuncakRinard03ExistentialHeapAbstractionEntailment,
    Yorsh03LogicalCharacterizationsHeapAbstractions}.
\end{example}

\newcommand{\FS}{\bigvee_{i=1}^n F_i}
\begin{figure}
  \begin{equation*}
    \begin{array}{rcl}
      \rtcop{F} & \equiv & \rtrancl\, (\lambda \lambda F)\, \ind{2}\, \ind{1} \mnl
      \tr{\rtrancl}\, r\, x\, y &=& 
      \exists n \geq 0.
      \exists z_0,\ldots,z_n.\
      z_0=x \land z_n=y \ \land \mnl
      & &
      \bigwedge_{i=0}^{n-1} r\, z_i\, z_{i+1} \mnl
      F_1 \circ F_2 & \equiv & 
      \curlyb{(\lambda \lambda F_1)\ind{3}\ind{1} \ \land \
              (\lambda \lambda F_1)\ind{2}\ind{1}} \mnls
      \tcop{F} & \equiv & F \circ \rtcop{F} \mnls
      \acyclic\ F & \equiv & \lnot \curlyb{\tcop{F} \land \ID} \mnls
      \tree\ F_1, \ldots, F_n
      & \equiv  & \acyclic\ \FS \mnls
      & \land & [\begin{array}[t]{@{}l}
        \rtcop{(\FS)} \implies \mnls
        \sum_{i=1}^n \Card\, (\twid{F_i}) \leq 1]
      \end{array}
    \end{array}
  \end{equation*}
  \caption{Transitive Closure Construct and Shorthands
    \label{fig:rtrancl}}
\end{figure}

For additional expressive power we introduce the
reflexive-transitive closure operator $*$, with the
semantics in Figure~\ref{fig:rtrancl}.  We also introduce
a shorthand for relation composition.  The relation
composition shorthand works when $F_1$ and $F_2$ both denote
binary relations, when the resulting expression can be
thought of as denoting a binary relation, as well as when
$F_1$ denotes a set and $F_2$ denotes a binary relation,
when the resulting expression denotes the set which is the
image of $F_1$ under $F_2$.  For the case of relation we
also introduce a simpler definition in
Figure~\ref{fig:rltwoShorthands} whose advantage is that it
uses only two implicit indices.



\subsection{Role Logic}
\label{sec:roleLogic}

\begin{figure*}
  \begin{equation*}
    \begin{array}{rclr}
      \Formula & = & 
      \Vars
      & \cmnt{named object or predicate} \mnls
      & \mid & \ind{\Nat}
      & \cmnt{de Bruijn index of an object variable} \mnls
      & \mid & \ID
      & \cmnt{equality between $\ind{1}$ and $\ind{2}$} \mnls
      & \mid & \Formula \land \Formula
      & \cmnt{conjunction} \mnls
      & \mid & \lnot \Formula 
      & \cmnt{negation} \mnls
      & \mid & \exists \Formula
      & \cmnt{existential quantification over objects} \mnls
      & \mid & \lambda \Formula
      & \cmnt{de Bruijn abstraction over objects} \mnls
      & \mid & \lambda \Vars : \Type\ .\ \Formula 
      & \cmnt{abstraction over named variables} \mnls
      & \mid & \Formula\ \Formula
      & \cmnt{function application} \mnls
      & \mid & \Formula'
      & \cmnt{let $\ind{1}$ be $\ind{2}$ in $F$} \mnls
      & \mid & \twid{\Formula}
      & \cmnt{relation inverse} \mnls
      & \mid & \card{{\geq}k} \Formula
      & \cmnt{at least $k$ objects satisfy $F$} \mnls
      & \mid & \Formula *
      & \cmnt{reflexive transitive closure}
    \end{array}
  \end{equation*}
  \caption{The Syntax of Role Logic\label{fig:RoleLogicSyntax}}
\end{figure*}

Figure~\ref{fig:RoleLogicSyntax} summarizes the syntax of
role logic.  The semantics of role logic follows from
Section~\ref{sec:recipe}.  

We next explain the purpose of lambda abstraction in our
logic.

\subsection{Lambda Calculus for Predicate Definitions}
\label{sec:lambdaDefinitions}

In the resulting role logic of
Figure~\ref{fig:RoleLogicSyntax} we retain the named
variables in the environment, and we allow abstraction over
those named variables.  As a result, there two kinds of
lambda abstraction: abstraction over de Bruijn indices and
abstraction over named variables.  Abstraction over a de
Bruijn index is always over $\ind{1}$ which denotes an
object of type $\obj$, such abstraction is written $\lambda
F$.  The abstraction over a named variable may abstract over
variables of more complex types and is written $\lambda
x:T.F$.  There is only one kind of lambda calculus
application; both $(\lambda F_1) F_2$ and $(\lambda x:T.F_1)
F_2$ are redexes.

The purpose of the named lambda abstraction $\lambda x:T.F$
is twofold.  First, when $T \equiv \obj$, then we can write
$\exists (\lambda x:\obj.F)$ as $\exists x.F$ as in the
usual first-order predicate calculus.  Second, when $T$ is
not $\obj$, we can encode acyclic definitions of
higher-order predicates that can be subsequently substituted
away.  Define the expression
\begin{equation*}
  \qlet\ P:T=F_1\ \qin\ F_2
\end{equation*}
to be equivalent to
\begin{equation*}
  (\lambda P:T\ .\ F_2) F_1
\end{equation*}
Such definitions are very useful for describing complex data
structures.

Note that acyclic definitions introduced through typed
lambda calculus via bindings $\lambda x:T.F$ for $T
\not\equiv \bool$ do not make the logic higher-order,
because we define the the quantifier $\exists$ to always
have the monomorphic type $(\obj \to \bool) \to \bool$,
and the reflexive-transitive closure operator $*$
to have the type
\begin{equation*}
  (\obj \to \obj \to \bool) \to (\obj \to \obj \to \bool)
\end{equation*}
Consider a well-typed formula $F$ whose only free
variables are relation symbols, and whose de Bruijn indices
only refer to indices bound in the formula.  Assume that we
have applied the Default Argument Rule, so that all de
Bruijn indices are explicit.  Then we may treat de Bruijn
abstraction as the usual abstraction over a disjoint set of
variables.  By strong normalization of simply typed lambda
calculus \cite{Barendregt01LambdaCalculiTypes}, let $F^0$ be
the normal form of $F$.  We claim that in $F^0$ the only
occurrence of lambda abstraction is within expressions of the
form $\exists (\lambda x:\obj.F)$ or $\rtrancl (\lambda
x:\obj.\lambda y:\obj.F)$.

To show the claim, consider an occurrence of $\lambda
x:\obj.F_0$ in $F^0$.  Let $F_1$ be the largest enclosing
occurrence $\lambda x_1:T_1. \ldots. \lambda x_n:T_n.\lambda
x:\obj.F_0$.  Then $F_1$ cannot be the entire $F^0$, because
$F^0$ has type $\bool$ by subject reduction.  $F_1$ cannot
occur within some application $F_1 F_2$, because $F_1 F_2$
would constitute a redex and $F^0$ is in normal form.  Hence,
$F_1$ can only occur in an expression of the form $F_3 F_1$.
Let us consider the ``spine''
\cite{Jones87ImplementationFunctionalProgrammingLanguages}
of $F_3 F_1$, so $F_3 \equiv F_n F_{n-1} \ldots F_4$ $n \geq
3$ and $F_n$ is not an application.  $F_n$ is not an
abstraction, because $F^0$ is in normal form.  Hence, $F_n$
can only be a variable or a constant.  

The only variables or or constants that can, by the typing
rules, be applied to an abstraction $F_1$ are $\exists$ and
$\rtrancl$, so either $F_n \equiv \exists$ or $F_n \equiv
\rtrancl$.

Consider the case $F_n \equiv \exists$.  By the type of
$\exists$, we conclude $F_3 \equiv F_n$ and $F_1 \equiv
\lambda x:\obj.F_0$, as desired.

Consider the case $F_n \equiv \rtrancl$.  Then $F_3 \equiv
F_n$, and $F_1 \equiv \lambda u:\obj.\lambda v:\obj.G$, so
either $u\equiv x$ and $F_1 \equiv \lambda x.\obj.F_0$ where
$F_0 = \lambda v:\obj.G$, or $v \equiv x$ and $F_1 \equiv
\lambda u:\obj.\lambda x:\obj.F_0$.  This finishes the proof
of the claim.

We conclude that typed lambda calculus allows us to use
flexible definitions of higher-order predicates to structure
our specifications while keeping the language first-order,
because we may substitute away all definitions using strong
normalization of the typed lambda calculus.



\section{Role Logic Subset $\RLtwo$ and its Decidability}
\label{sec:decidable}

In this section we introduce a subset $\RLtwo$ of role logic
(Figure~\ref{fig:RLtwoSyntax}) and show its decidability.

To show the decidability of $\RLtwo$, we give translations
of formulas between the following four logics:
\begin{enumerate}
\item $D^2$: the formulas of the first-order logic with counting
  in which every subformula has at most two
  free variables (different subformulas may have different 
  free variables);
\item $C^2$: the formulas of the two-variable logic with
  counting, which uses $x$ and $y$ as the only variable
  names; the satisfiability and finite satisfiability
  problem for $C^2$ was shown to be decidable in
  \cite{GraedelETAL97TwoVariableLogicCountingDecidable}; the
  satisfiability problem for $C^2$ was shown
  $\NEXPTIME$-complete in
  \cite{PacholskiETAL00ComplexityResultsFirstOrderTwo};
\item $I^2$: de Bruijn version of the two-variable logic
  with counting, which uses only de Bruijn indices $\ind{1}$
  and $\ind{2}$;
\item $\RLtwo$: a subset of role logic that contains no
  explicit de Bruijn indices.
\end{enumerate}

\begin{figure}
  \begin{center}
    {\large
\setlength{\unitlength}{0.42in}
\begin{picture}(4,4)
\put(1,1){\makebox(0.5,0.5){$\RLtwo$}}
\put(3,1){\makebox(0.5,0.5){$I^2$}}
\put(1,3){\makebox(0.5,0.5){$D^2$}}
\put(3,3){\makebox(0.5,0.5){$C^2$}}
\put(2.8,1.25){\vector(-1,0){1.1}}
\put(1.7,3.25){\vector(1,0){1.1}}
\put(1.25,1.5){\vector(0,1){1.5}}
\put(3.25,3.0){\vector(0,-1){1.5}}
\put(1.8,3.25){\makebox(1,0.5){\normalsize Fig.~\ref{fig:DtoC}}}
\put(1.8,1.25){\makebox(1,0.5){\normalsize Fig.~\ref{fig:itwoToRLtwo}}}
\put(0.15,1.9){\makebox(1,0.5){\normalsize Fig.~\ref{fig:RLtwoToDtwo}}}
\put(3.4,1.9){\makebox(1,0.5){\normalsize Fig.~\ref{fig:CtwoToItwo}}}
\end{picture}
}
%
  \end{center}  
  \caption{Showing Equivalence of Four Logics.
    \label{fig:showingEquivalence}}
\end{figure}
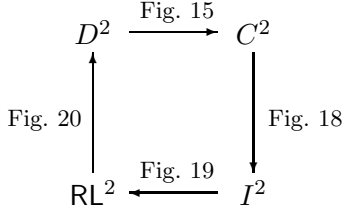
Figure~\ref{fig:showingEquivalence} sketches the idea of the
proof of equivalence of these four logics.  We give
translations of formulas from $D^2$ to $C^2$
(Section~\ref{sec:twoVarLogic}, Figure~\ref{fig:DtoC}) from
$C^2$ to $I^2$ (Section~\ref{sec:itwoConnection},
Figure~\ref{fig:CtwoToItwo}), from $I^2$ to $\RLtwo$
(Section~\ref{sec:itwoConnection}, Figure~\ref{fig:itwoToRLtwo})
and from $\RLtwo$ to $D^2$ (Section~\ref{sec:RLtoDtwo},
Figure~\ref{fig:RLtwoToDtwo}).  These translations imply
that the satisfiability problem for these four logics are
equivalent, so by decidability of $C^2$
\cite{GraedelETAL97TwoVariableLogicCountingDecidable} we
conclude that all these logics are decidable.



  \subsection{The Role Logic Subset $\RLtwo$}

\begin{figure}
  \begin{equation*}
    \begin{array}{rclr}
      \Formula & = & 
      \Vars
      & \cmnt{binary or unary relation symbol} \mnls
      & \mid & \ID
      & \cmnt{equality between $\ind{1}$ and $\ind{2}$} \mnls
      & \mid & \Formula \land \Formula
      & \cmnt{conjunction} \mnls
      & \mid & \lnot \Formula 
      & \cmnt{negation} \mnls
      & \mid & \Formula'
      & \cmnt{let $\ind{1}$ be $\ind{2}$ in $F$} \mnls
      & \mid & \twid{\Formula}
      & \cmnt{relation inverse} \mnls
      & \mid & \card{{\geq}k} \Formula
      & \cmnt{at least $k$ objects satisfy $F$} \mnls
    \end{array}
  \end{equation*}
  \caption{The Syntax of $\RLtwo$ Subset of Role Logic\label{fig:RLtwoSyntax}}
\end{figure}

\begin{figure}
  \begin{equation*}
    \begin{array}{rcl}
      \Nat_2 & = & \{ 1, 2 \} \mnl
      e & :: & \Nat_2 \to \obj \mnl
      \tr{A}e & = & \tr{A}(e\, 1) \mnl
      \tr{f}e & = & \tr{f}(e\, 2, e\, 1) \mnl
      \tr{\ID}e & = & (e\, 2) = (e\, 1) \mnl
      \tr{F_1 \land F_2}e & = & (\tr{F_1}e) \land (\tr{F_2}e) \mnl
      \tr{\lnot F}e & = & \lnot (\tr{F}e) \mnl
      \tr{F'}e & = & \tr{F} (e[1 \mapsto (e\, 2)]) \mnl
      \tr{\twid{F}}e & = & \tr{F} (e[1 \mapsto (e\, 2), 2 \mapsto (e\, 1)]) \mnl
      \tr{\card{{\geq}k} F}e & = & | \{ o \mid \tr{F} (e[1 \mapsto o, 2 \mapsto (e\, 1)]) \} | \geq k
    \end{array}
  \end{equation*}
  \caption{The Semantics of $\RLtwo$\label{fig:RLtwoSemantics}}
\end{figure}

\begin{figure}
  \begin{equation*}
    \begin{array}{rcl}      
      \multicolumn{3}{l}{\mbox{quantifiers:}} \mnl
      \curlyb{F} & = & \card{{\geq}1} F \mnl
      \squareb{F} & = & \lnot \curlyb{\lnot F} \mnl      
      \multicolumn{3}{l}{\mbox{relation image:}} \mnl
      \image{F_A}{F_r} & = & \curlyb{F_A \land \twid{F_r}} \mnl
      \multicolumn{3}{l}{\mbox{weakest precondition:}} \mnl
      \wlp{F_r}{F_A} & = & \squareb{F_r \implies F_A}
    \end{array}
  \end{equation*}
  \caption{Some Shorthands for $\RLtwo$\label{fig:rltwoShorthands}}
\end{figure}

Figure~\ref{fig:RLtwoSyntax} presents the two-variable role
logic $\RLtwo$.  Compared to the full role logic in
Figure~\ref{fig:RoleLogicSyntax}, $\RLtwo$ omits the
constructs for creating definitions, the constructs for
explicitly referring to object variables, and transitive
closure.  Figure~\ref{fig:RLtwoSemantics} summarizes the
semantics of $\RLtwo$; this semantics is in accordance with
the semantics of the full role logic derived in
Section~\ref{sec:recipe}.  Figure~\ref{fig:rltwoShorthands}
defines shorthands that illustrate some constructs definable
in $\RLtwo$.

We show that $\RLtwo$ has precisely the same expressive
power as the set of the formulas of logic $C^2$, which is
shown decidable in
\cite{GraedelETAL97TwoVariableLogicCountingDecidable} over
the set of all models, as well as over the set of finite
models.


  \subsection{Two-Variable Logics $C^2$ and $D^2$}
\label{sec:twoVarLogic}

\begin{figure*}
  \begin{equation*}
    \begin{array}{rclr}
      \Vars_2 & = & \{ x, y \} \mnl
      \Formula & = & A(\Vars_2)
      & \cmnt{atomic formula with unary relation $A$} \mnls
      & \mid & f(\Vars_2, \Vars_2)
      & \cmnt{atomic formula with binary relation $f$} \mnls
      & \mid & \Vars_2 = \Vars_2
      & \cmnt{equality between objects} \mnls
      & \mid & \Formula \land \Formula
      & \cmnt{conjunction} \mnls
      & \mid & \lnot \Formula 
      & \cmnt{negation} \mnls
      & \mid & \existsgeq{k}{\Vars_2}{\Formula}
      & \cmnt{at least $k$ objects satisfy formula} \mnls
    \end{array}
  \end{equation*}
  \caption{The Syntax of Two-Variable Logic with Counting $C^2$\label{fig:twoVarSyntax}}
\end{figure*}

Figure~\ref{fig:twoVarSyntax} presents the logic $C^2$
\cite{GraedelETAL97TwoVariableLogicCountingDecidable}.  The
logic $C^2$ is first-order logic with equality and counting,
restricted to formulas that contain only two fixed
variable names $x$ and $y$.  

In this section we argue that a more flexible restriction on
variable names yields logic with same definable relations.
Let $\FV(F)$ denote the free variables of formula $F$.

\begin{definition} \label{def:dtwo}
  A $D^2$ formula is a formula $F$ of first-order logic with
  counting such that $|\FV(G)| \leq 2$ for every subformula
  $G$ of $F$.
\end{definition}

Clearly every $C^2$ formula is a $D^2$ formula, but not vice
versa, because the set of possible variables that may occur
in $D^2$ formulas is countably infinite.  The syntactic
restriction on variables in Definition~\ref{def:dtwo} is
more general than in the definition in $C^2$, which makes
$D^2$ more convenient for writing readable formulas.

We show that every $D^2$ formula is equivalent to a $C^2$
formula (modulo the renaming of free variables).  Up to one
technical detail, it suffices to rename bound variables in a
$D^2$ formula to obtain a $C^2$ formula.  We therefore
derive the equivalence of $D^2$ and $C^2$ as a consequence
of an observation about lambda calculus terms.

\begin{definition} \label{def:niceTerms}
  Define the set of lambda calculus terms $\NiceTerms$ as
  the smallest set that satisfies the following conditions:
  \begin{enumerate}
  \item $v \in \NiceTerms$ if $v$ is a variable and
    $c \in \NiceTerms$ if $c$ is a constant;
  \item if $T_1, T_2 \in \NiceTerms$ and
    $|\FV(T_1) \cup \FV(T_2)| \leq 2$, then 
    $(T_1 T_2) \in \NiceTerms$;
  \item if $T \in \NiceTerms$, $v$ is a variable,
    and $|\FV(T) \cup \{ v \}| \leq 2$, then
    $\lambda v.T \in \NiceTerms$.
  \end{enumerate}
\end{definition}
From Definition~\ref{def:niceTerms} it follows that if $T
\in \NiceTerms$, then $|\FV(T_1)|\leq 2$ for every subterm
$T_1$ of $T$.  Moreover, if $\lambda v.T \in \NiceTerms$ and
$v \notin \FV(T)$, then $|\FV(T)| \leq 1$.

We next define the set $\capturing{v}{F}$ of those bound
variables $z$ in formula $F$ such that $v$ occurs in the
scope of a binding of $z$.
\begin{definition}
  \begin{equation*}
    \begin{array}{rcl}
      \capturing{v}{u} & = & \emptyset, \mbox{ if $u$ is a variable} \mnl
      \capturing{v}{F_1 F_2} & = & \capturing{v}{F_1} \cup \capturing{v}{F_2} \mnl
      \capturing{v}{\lambda u.F} & = & 
      \left\{ \begin{array}{rl}
          \capturing{v}{F} \cup \{ u \}, & \mbox{ if } v \in \FV(\lambda u.F) \mnl
          \emptyset, & \mbox{ otherwise }
        \end{array}\right.
    \end{array}
  \end{equation*}
\end{definition}

As usual, we say that $T$ and $T'$ are $\alpha$-equivalent
if $T'$ can be obtained from $T$ by renaming bound
variables.
\begin{lemma} \label{lemma:renaming}
  For every $T \in \NiceTerms$ with $\FV(T) \subseteq \{ u,v
  \}$ there exists a term $T' = \nort{T}$ such that $T'$ is
  $\alpha$-equivalent to $T$, all bound variables in $T'$
  are among $\{x, y\}$, and either
  \begin{enumerate}
  \item $\capturing{u}{T'} \subseteq \{x\}$ and
    $\capturing{v}{T'} \subseteq \{y\}$, or
  \item $\capturing{u}{T'} \subseteq \{y\}$ and
    $\capturing{v}{T'} \subseteq \{x\}$.
  \end{enumerate}
\end{lemma}
\begin{proof}
  Let $\FV(T) \subseteq \{ u, v \}$.  Without loss of
  generality we may assume that $\{ u, v \} \cap \{ x, y \}
  = \emptyset$.  The proof is by induction on the structure
  of terms.
  \begin{enumerate}
  \item $T = u$ for a variable $u$.  Let $T = T'$,
    clearly $\capturing{u}{T'} = \capturing{v}{T'} = \emptyset$.
  \item $T = T_1 T_2$.  Let $T'_1 = \nort{T_1}$ and
    $T'_2 = \nort{T_2}$ by induction hypothesis.
    Assume $\capturing{u}{T'_1} \subseteq \{ x \}$
    and $\capturing{v}{T'_1} \subseteq \{ y \}$
    (the other case is symmetric).  We consider
    two cases for $T'_2$.
    \begin{enumerate}
    \item $\capturing{u}{T'_2} \subseteq \{ x \}$
      and $\capturing{v}{T'_2} \subseteq \{ y \}$.
      Then let $\nort{T} = T'_1 T'_2$.
    \item $\capturing{u}{T'_2} \subseteq \{ y \}$ and
      $\capturing{v}{T'_2} \subseteq \{ x \}$.  Let $T''_2$
      be the result of swapping in $T'_2$ all occurrences of
      bound variables $x$ and $y$.  Then 
      $\capturing{u}{T''_2} \subseteq \{ x \}$
      and $\capturing{v}{T''_2} \subseteq \{ y \}$,
      so we let $\nort{T} = T'_1 T''_2$.
    \end{enumerate}
    In both cases, $\capturing{u}{\nort{T}} \subseteq \{ x
    \}$ and $\capturing{v}{\nort{T}} \subseteq \{ y \}$.
  \item $T = \lambda w. T_1$.  $|\{ u, v \}|=2$ and
    $|\FV(T_1) \cup \{ w \}|\leq 2$ by the definition of
    $\NiceTerms$, so it cannot be the case that both $u \in
    \FV(T_1)$ and $v \in \FV(T_1)$.  Since $\FV(T_1)
    \subseteq \{ u, v, w \}$, we conclude that $\FV(T_1)
    \subseteq \{ u, w \}$ or $\FV(T_1) \subseteq \{ v, w
    \}$.
    
    Suppose therefore that $\FV(T_1) \subseteq \{ u, w \}$
    (the case $\FV(T_1) \subseteq \{ v, w \}$ is symmetric).
    By induction hypothesis, let $T'_1 = \nort{T_1}$.
    Assume $\capturing{u}{T_1} \subseteq \{ x \}$ and
    $\capturing{w}{T_1} \subseteq \{ y \}$ (the case
    $\capturing{u}{T_1} \subseteq \{ y \}$ and
    $\capturing{w}{T_1} \subseteq \{ x \}$ is symmetric).
    Let $\nort{T} = \lambda x. (F_1[w:=x])$.  Then
    $\capturing{u}{\nort{T}} \subseteq \{x\}$ and
    $\capturing{v}{\nort{T}} = \emptyset \subseteq \{ y \}$.
  \end{enumerate}
\end{proof}

To apply Lemma~\ref{lemma:renaming} to $D^2$ formulas, we
represent all logical operations and quantifiers as
constants.  Variables in a lambda term then correspond to
first-order variables.  To ensure that the representation of
formulas satisfies the condition $|\FV(T) \cup \{ v \}| \leq
2$ for each term $\lambda v.T$, we require the following
condition:
\begin{equation} \label{eqn:formulaCondition}
  \begin{array}{c}
    \mbox{For every formula $\existsgeq{k}{x}{F}$,} \\ 
    \mbox{either $x \in \FV(F)$ or $F \equiv \boolTrue$.}
  \end{array}
\end{equation}
We ensure this condition by applying the rule
\begin{equation*}
  \existsgeq{k}{x}{F} \ssim F \land \existsgeq{k}{x}{\boolTrue}
\end{equation*}
for $x \notin \FV(F)$.  

\begin{figure*}
  \begin{equation*}
    \begin{array}{rcl}
      \dctr{A(v)} & = & A(v) \mnl
      \dctr{f(u,v)} & = & f(u,v) \mnl
      \dctr{\lnot F} & = & \lnot \dctr{F} \mnl
      \dctr{F_1 \land F_2} & = & 
      \left\{\begin{array}{cl}
          F'_1 \land F'_2, & \mbox{ if }
          \begin{array}[t]{@{}l}
            \capturing{u}{F'_1}, \capturing{u}{F'_2} \subseteq \{ x \} \\
            \capturing{v}{F'_1}, \capturing{v}{F'_2} \subseteq \{ y \} \\
            \mbox{ or } \\
            \capturing{u}{F'_1}, \capturing{u}{F'_2} \subseteq \{ y \} \\
            \capturing{v}{F'_1}, \capturing{v}{F'_2} \subseteq \{ x \} \mnl
          \end{array} \\
          F'_1 \land (\swap\ F'_2), & \mbox{ otherwise} \mnl
        \end{array}\right. \\
      \\
      & & \begin{array}[t]{l}
        \FV(F_1 \land F_2) = \{ u, v \} \mnl
        F'_1 = \dctr{F_1} \mnl
        F'_2 = \dctr{F_2} \\
        \\
        \begin{array}[t]{@{}lcl}
          \swap\, (A(v)) & = & A(s\, u, s\, v) \mnl
          \swap\, (f(u,v)) & = & f(s\, u, s\, v) \mnl
          \swap\, (\lnot F) & = & \lnot (\swap\, F) \mnl
          \swap\, (F_1 \land F_2) & = & \swap\, F_1 \ \land\ \swap\, F_2 \mnl
          \swap\, (\existsgeq{k}{v}{F}) & = &
          \existsgeq{k}{(s\, v)}{(\swap\, F)} \mnl
          \multicolumn{3}{l}{
          \begin{array}[t]{l}
            s\, x = y, \quad
            s\, y = x \\
            s\, u = u, \mbox{ if } u \notin \{ x, y \} \mnl
          \end{array}}
        \end{array}
      \end{array} \\
      \dctr{\existsgeq{k}{w}{F}} & = &
      \left\{\begin{array}{cl}
          \existsgeq{k}{x}{(F'[w:=x])}, & 
          \mbox{ if }
          \capturing{u}{F'} \subseteq \{ x \},
          \capturing{w}{F'} \subseteq \{ y \} \mnl
          \existsgeq{k}{y}{(F'[w:=y])}, & 
          \mbox{ if }
          \capturing{u}{F'} \subseteq \{ y \},
          \capturing{w}{F'} \subseteq \{ x \} \\
        \end{array}\right. \\
      \\
      & & \begin{array}[t]{l}
        \FV(F) \subseteq \{ u, w \} \mnl
        F' = \dctr{F}
      \end{array}
    \end{array}
  \end{equation*}
  \caption{Translation of $D^2$ formulas to $C^2$ formulas.
    \label{fig:DtoC}}
\end{figure*}

After ensuring the condition~(\ref{eqn:formulaCondition}),
we apply the translation in Figure~\ref{fig:DtoC}.
Lemma~\ref{lemma:renaming} justifies the correctness of the
translation.  The translated formula is of the same size as
the original formula.  The translation can clearly be
performed in polynomial time, including the process of
ensuring the condition~(\ref{eqn:formulaCondition}).  The
translation time can be made close to linear by delaying the
application of the substitution $[w:=x]$ and the $\swap$
operation.



  \subsection{From $C^2$ to $\RLtwo$ via $I^2$}
\label{sec:itwoConnection}

In this section we introduce logic $I^2$
(Figure~\ref{fig:itwoSyntax}).  We then give translations
from $C^2$ to $I^2$ (Figure~\ref{fig:CtwoToItwo}), and from
$I^2$ to $\RLtwo$ (Figure~\ref{fig:itwoToRLtwo}).

\begin{figure*}
  \begin{equation*}
    \begin{array}{rclr}
      \Formula & = & A(\ind{\Nat_2})
      & \cmnt{atomic formula with unary relation $A$} \mnls
      & \mid & f(\ind{\Nat_2}, \ind{\Nat_2})
      & \cmnt{atomic formula with binary relation $f$} \mnls
      & \mid & \ind{\Nat_2} = \ind{\Vars_2}
      & \cmnt{equality between objects} \mnls
      & \mid & \Formula \land \Formula
      & \cmnt{conjunction} \mnls
      & \mid & \lnot \Formula 
      & \cmnt{negation} \mnls
      & \mid & \card{{\geq}k}\Formula
      & \cmnt{at least $k$ objects satisfy formula} \mnls
    \end{array}
  \end{equation*}
  \caption{The Syntax of Intermediate Logic $I^2$\label{fig:itwoSyntax}}
\end{figure*}

\paragraph{Intermediate logic.}
Figure~\ref{fig:itwoSyntax} presents logic $I^2$.  $I^2$ is
a version of $C^2$ that uses two de Bruijn indices instead
of variables.  We introduce $I^2$ to separate the the
translation of $C^2$ formulas to $\RLtwo$ in two phases: the
first phase introduces de Bruijn indices, and the second
phase introduces Default Argument Rule.

For the sake of illustration, we first present a converse
translation, from $I^2$ to $C^2$, although we do not need
this translation to show the equivalence of $D^2$, $C^2$,
$I^2$, and $\RLtwo$.

\begin{figure}
  \begin{equation*}
    \begin{array}{rcl}
      e & :: & \Nat_2 \to \Vars_2 \mnl
      \ictr{A(\ind{i})}e & = & A(e\, i) \mnl
      \ictr{f(\ind{i_1},\ind{i_2})}e & = & f(e\, i_1, e\, i_2) \mnl
      \ictr{\ind{i_1} {=} \ind{i_2}}e & = & (e\, i_1) = (e\, i_2) \mnl
      \ictr{F_1 \land F_2}e & = & (\ictr{F_1}e) \land (\ictr{F_2}e) \mnl
      \ictr{\lnot F}e & = & \lnot (\ictr{F}e) \mnl
      \ictr{\card{{\geq}k}F}e & = & 
        \begin{array}[t]{@{}l}
          \existsgeq{k}{v}{(\ictr{F}[1 \mapsto v, 2 \mapsto (e\, 1)])} \\
          v = s(e\, 1) \mnl
          s\, x = y, \quad s\, y = x \\
        \end{array} \\
      \\
      \multicolumn{3}{c}{\mbox{\bf correctness criterion:}} \mnl
      \multicolumn{3}{c}{\tr{\ictr{F}e}e_C = \tr{F} (e_C \circ e)}
    \end{array}
  \end{equation*}
  \caption{Translating $I^2$ formulas to $C^2$ formulas\label{fig:itwoToCtwo}}
\end{figure}

\paragraph{From $I^2$ to $C^2$.}
Figure~\ref{fig:itwoToCtwo} presents the translation of
$I^2$ into $C^2$.  This translation amounts to introducing
alternatively variables $x$ and $y$ for each counting
quantifier, and resolving the indices appropriately.  Using
the criterion in Figure~\ref{fig:itwoToCtwo}, the
correctness of the translation follows by induction on the
structure of formulas.

\paragraph{From $C^2$ to $I^2$.}
We turn to the translation from $C^2$ to $I^2$.
Consider the $C^2$ formula 
\begin{equation*}
  F \ \equiv \
  \existsgeq{1}{y}{(\existsgeq{1}{x}{(\existsgeq{1}{x}{P(x,y)}) \land Q(x,y)})}
\end{equation*}
The subformula $P(x,y)$ of $F$ refers to the variable $y$,
which is the 3rd bound variable starting from the innermost
one.  Therefore, the straightforward replacement of
variables by de Bruijn indices would require the access to $\ind{3}$.
To address this problem, the translation from $C^2$ to $I^2$
involves a preparatory ``alternating transformation'' on
$C^2$ formulas.  For every formula $F$, let $B(F)$ denote
some purely propositional combination of $F$ and perhaps
some other formulas.  The alternating transformation
eliminates all subformulas of the form
$\existsgeq{k_1}{v}{B(\existsgeq{k_2}{v}{G(v)})}$ for $v \in
\Vars_2$.  In the resulting formula, the sequence of bound
variables along any path in the formula tree is alternating,
that is, satisfies the regular expression $(y |
\epsilon)(xy)^{*}(x | \epsilon)$.

For the purpose of alternating transformation, we add the
disjunction $\lor$ to the language.  We show how to
eliminate successive quantification over $x$ from
$\existsgeq{k_1}{x}{B(\existsgeq{k_2}{x}{G})}$ (the case of
$\existsgeq{k_1}{y}{B(\existsgeq{k_2}{y}{G})}$ is
analogous).  First, transform $B$ into disjunction of
canonical conjunctions of formulas $H$, where each $H$
satisfies one of the following three conditions:
\begin{enumerate}
\item[$C1)$] $H$ is quantifier-free;
\item[$C2)$] $H$ is of the form $\existsgeq{k}{v}{G(v)}$
  for $v \in \Vars_2$;
\item[$C2)$] $H$ is of the form $\lnot \existsgeq{k}{v}{G(v)}$
  for $v \in \Vars_2$;
\end{enumerate}
Let $B \equiv \bigvee_{i=1}^n B_i$ where each $B_i$ is a
canonical conjunction (cube) of formulas satisfying
conditions $C1)$, $C2)$, $C3)$.  Because $B_i \land B_j$ is
contradictory for distinct cubes $B_i$ and $B_j$, the sets
of objects $o$ satisfying different $B_i$ are disjoint, so
\begin{equation*}
  | \{ o \mid \tr{B}e[v \to o] \} | =
  \sum_{i=1}^n | \{ o \mid \tr{B_i}e[v \to o] \} |
\end{equation*}
We can therefore replace counting quantifier on $B$
with a propositional combination of counting quantifiers
on $B_i$ for $1 \leq i \leq n$ (as in quantifier
elimination for boolean algebras, \cite{Skolem19Untersuchungen},
\cite[Section 3.2]{KuncakRinard03TheoryStructuralSubtyping}).
Specifically,
\begin{equation} \label{eqn:splittingCard}
 \existsgeq{k_1}{x}{B} \ssim
   \bigvee_{\sum_{j=1}^n l_j = k_1} \ \
     \bigwedge_{i=1}^n \existsgeq{l_i}{x}{B_i}
\end{equation}
It is therefore sufficient to eliminate the successive
quantification over $x$ in
$\existsgeq{k_1}{x}{B_i(\existsgeq{k_2}{x}{G})}$.  Group the
conjuncts in $B_i$ as follows.  Let $\FV(F)$ denote free
variables of formula $F$.  Let $P(x)$ be the conjunction
of conjuncts $C$ of $B_i$ such that $x \in \FV(C)$, and let
$Q$ be the conjunction of all conjuncts $C$ of $B_i$ such
that $x \notin \FV(C)$.  All occurrences of
$\existsgeq{k_2}{x}{G}$ in $B_i$ are in $Q$.  We have
\begin{equation*}
  \begin{array}{l}
    \existsgeq{k_1}{x}{B_i} \ssim
    \existsgeq{k_1}{x}{Q \land P(x)} \ssim
    Q \land \existsgeq{k_1}{x}{P(x)}
  \end{array}
\end{equation*}
where the last equivalence follows easily by definition of
the counting quantifier $\existsgeq{x}{k_1}{}$  In the
resulting formula $Q \land \existsgeq{x}{k_1}{P(x)}$, the
subformula $\existsgeq{k_2}{x}{G}$ is in $Q$ and is
therefore not in the scope of the original quantifier.  By
repeating this transformation we ensure that all quantifiers
are alternating.

\begin{figure}
  \begin{equation*}
    \begin{array}{rcl}
      e & :: & \Vars_2 \to \Nat_2 \mnl
      \citr{A(v)}e & = & A(\ind{e\, v}) \mnl
      \citr{f(v_1,v_2)}e & = & f(\ind{e\, i_1}, \ind{e\, i_2}) \mnl
      \citr{v_1 {=} v_2}e & = & \ind{e\, v_1} = \ind{e\, i_2} \mnl
      \citr{F_1 \land F_2}e & = & (\citr{F_1}e) \land (\citr{F_2}e) \mnl
      \citr{\lnot F}e & = & \lnot (\citr{F}e) \mnl
      \citr{\existsgeq{k}{x}{F}}e & = &
        \begin{array}[t]{@{}l}
          \card{{\geq}k} (\citr{F}[x \mapsto 1, y \mapsto 2]) \mnls
          \mbox{invariant:}\ \ e\, y = 1 \mnl
        \end{array} \mnl
      \citr{\existsgeq{k}{y}{F}}e & = &
        \begin{array}[t]{@{}l}
          \card{{\geq}k} (\citr{F}[y \mapsto 1, x \mapsto 2]) \mnls
          \mbox{invariant:}\ \ e\, x = 1 \mnl
        \end{array} \\
      \\
      \multicolumn{3}{c}{\mbox{\bf correctness criterion:}} \mnl
      \multicolumn{3}{c}{\tr{\citr{F}e}e_I = \tr{F} (e_I \circ e)}
    \end{array}
  \end{equation*}
  \caption{Translating normalized $C^2$ formulas to $I^2$ formulas\label{fig:CtwoToItwo}}
\end{figure}

After the alternating transformation, the translation from
$C^2$ to $I^2$ is straightforward, and is presented in
Figure~\ref{fig:CtwoToItwo}.  The correctness of the
translation follows by induction of the structure of
formulas.  The translation in Figure~\ref{fig:CtwoToItwo}
runs in linear time and produces an $I^2$ formula whose size
is linear in the size of the original $C^2$ formula.  

The alternating transformation that precedes the translation
may cause exponential blowup of the formula size due to
translation to disjunctive normal form, but for most
formulas the transformation need not be applied.  Moreover,
if we allow introducing new predicate names, then we may
replace $\existsgeq{k_1}{x}{B(\existsgeq{k_2}{x}{G(x,y)})}$
with $\existsgeq{k_1}{x}{B(P(y))}$ and conjoin the topmost
formula with the formula $\forall y.P(y) \iff
\existsgeq{k_2}{x}{G(x,y)}$.  Such transformation can be
performed in linear time and preserves the satisfiability of
formulas (see \cite[Section 2.1, Page
18]{GraedelETAL97TwoVariableLogicCountingDecidable} and
\cite[Lemma
2.3]{GraedelETAL97TwoVariableLogicCountingDecidable}).

\begin{figure}
  \begin{equation*}
    \begin{array}{rcl}
      \irtr{A(\ind{1})} & = & A \mnl
      \irtr{A(\ind{2})} & = & A' \mnl
      \irtr{f(\ind{2},\ind{1})} & = & f \mnl
      \irtr{f(\ind{1},\ind{2})} & = & \twid{f} \mnl
      \irtr{f(\ind{2},\ind{2})} & = & f' \mnl
      \irtr{f(\ind{1},\ind{1})} & = & \twid{(f')} \mnl
      \irtr{\ind{2}=\ind{1}} & = & \ID \mnl
      \irtr{\ind{1}=\ind{2}} & = & \ID \mnl
      \irtr{\ind{1}=\ind{1}} & = & \boolTrue \mnl
      \irtr{\ind{2}=\ind{2}} & = & \boolTrue \mnl
      \irtr{F_1 \land F_2} & = & \irtr{F_1} \land \irtr{F_2} \mnl
      \irtr{\lnot F} & = & \lnot \irtr{F} \mnl
      \irtr{\card{{\geq}k} F} & = & \card{{\geq}k} \irtr{F} \\
      \\
      \multicolumn{3}{c}{\mbox{\bf correctness criterion:}} \mnl
      \multicolumn{3}{c}{\tr{\irtr{F}}e_I = \tr{F}e_I}
    \end{array}
  \end{equation*}
  \caption{Translating $I^2$ formulas to $\RLtwo$ formulas\label{fig:itwoToRLtwo}}
\end{figure}

\paragraph{From $I^2$ to $\RLtwo$.}
Figure~\ref{fig:itwoToRLtwo} presents the translation from
$I^2$ to $\RLtwo$, which is simple and does not require a
translation environment.  The translation algorithm runs in
linear time and produces a $\RLtwo$ formula whose size is
linear in the size of the original $I^2$ formula.



  \subsection{From $\RLtwo$ to $D^2$: Closing the Loop}
\label{sec:RLtoDtwo}

\begin{figure}
  \begin{equation*}
    \begin{array}{rcl}
      e\, 0 & \in & \Nat \mnl
      e\, k & \in & \{ y_1, y_2, \ldots \}\ \ \mbox{ for } k \in \{ 1, 2 \}  \mnl
      \rdtr{A}e & = & A(e\, 1) \mnl
      \rdtr{f}e & = & f(e\, 2, e\, 1) \mnl
      \rdtr{\ID}e & = & (e\, 2) = (e\, 1) \mnl
      \rdtr{F_1 \land F_2}e & = & (\rdtr{F_1}e) \land (\rdtr{F_2}e) \mnl
      \rdtr{\lnot F}e & = & \lnot (\rdtr{F}e) \mnl
      \rdtr{\card{{\geq}k}F}e & = &
      \begin{array}[t]{@{}l}
        \existsgeq{k}{v}{\tr{F}e[0 \mapsto n, 1 \mapsto v, 2 \mapsto (e\, 1)]} \\
        v = y_n \\
        n = 1 + e\, 0 \mnl
      \end{array} \\
      \rdtr{\twid{F}}e & = & \rdtr{F}(e[1 \mapsto (e\, 2), 2 \mapsto (e\, 1)]) \mnl
      \rdtr{F'}e & = & \rdtr{F} (e[1 \mapsto (e\, 2)]) \\
      \\
      \multicolumn{3}{c}{\mbox{\bf correctness criterion:}} \mnl
      \multicolumn{3}{c}{\tr{\rdtr{F}e}e_C = \tr{F}(e_C \circ e)} \mnl
      \multicolumn{3}{c}{\mbox{\bf result is in $D^2$:}} \mnl
      \multicolumn{3}{c}{\FV(\rdtr{F}e) \subseteq \{ e\, 1, e\, 2 \}}
    \end{array}
  \end{equation*}
  \caption{Translating $\RLtwo$ formulas to $D^2$ formulas.\label{fig:RLtwoToDtwo}}
\end{figure}

In the final step, we provide a translation from $\RLtwo$ formulas
to $D^2$ formulas.  The logic $D^2$ is a convenient target of
translation of $\RLtwo$ formulas.
(Namely, a simple attempt at translation from $\RLtwo$ to $I^2$ runs
into the difficulty of the following form.  Formula
$(\card{{\geq}1} f)'$ is equivalent to $\card{{\geq}1}
f(\ind{3},\ind{1})$ which uses index $\ind{3}$ not available
in $I^2$.  Similarly, an attempt to translate from $\RLtwo$ to
$C^2$ runs into difficulty of variable capture.)

Figure~\ref{fig:RLtwoToDtwo} presents the
translation from $\RLtwo$ to $D^2$.  The correctness
of the translation follows by induction on the structure of formulas.
Furthermore, each subformula $G_1$ of a formula $\rdtr{F}e$
is of the form $G_1 \equiv \rdtr{G}e_1$ for some $G$ and
$r_1$, and by induction it follows that the free variables
of $\rdtr{G}e_1$ are among $\{ e_1\, 1, e_1\, 2 \}$.
Therefore, $|\FV(G_1)| \leq 2$ and the result of translation
is a $D^2$ formula.

\paragraph{Summary}
As indicated in Figure~\ref{fig:showingEquivalence}, we have
presented translations from $D^2$ to $C^2$, from $C^2$ to
$I^2$, from $I^2$ to $\RLtwo$, and from $\RLtwo$ to $D^2$.
We conclude that $D^2$, $C^2$, $I^2$, and $\RLtwo$ are all
equivalent logics, and, by
\cite{GraedelETAL97TwoVariableLogicCountingDecidable},
decidable.

The satisfiability problem for $C^2$ formulas is shown to be
$\NEXPTIME$-complete in
\cite{PacholskiETAL00ComplexityResultsFirstOrderTwo}.  We
have observed that there are efficient polynomial
transformations of formulas from $D^2$ to $C^2$, from $C^2$
to $I^2$, from $I^2$ to $\RLtwo$ and from $\RLtwo$ to $D^2$
that yield formulas equivalent for satisfiability.
(Moreover, all transformations except from $C^2$ to $I^2$
yield equivalent formulas in the same vocabulary.)  As a
result, the satisfiability problem of all these logics is
$\NEXPTIME$-complete.



\section{Applications of Role Logic}

We next present three applications of role logic.  In
Section~\ref{sec:transductions} we present a shape analysis
technique based on generating verification conditions in
$\RLtwo$ and applying the decision procedure for $\RLtwo$.
In Section~\ref{sec:bsac} we note that boolean shape
analysis constraints
\cite{KuncakRinard03OnBooleanAlgebraShapeAnalysisConstraints}
are a subset of constraints expressible in role logic.  In
Section~\ref{sec:dlencoding} we show that a different subset
of $\RLtwo$ corresponds to an expressive description logic
\cite[Chapter 5]{BaaderETAL03DescriptionLogicHandbook}.


  \subsection{Static Analysis Based on $\RLtwo$}
\label{sec:transductions}

This section shows how to use the decidability of $\RLtwo$
for static analysis of imperative programs.
Figure~\ref{fig:imperativeLanguage} presents the syntax of a
simple imperative language.
Figure~\ref{fig:statementPredicates} presents predicates in
$\RLtwo$ that describe the meaning of statements in this
language.

\begin{figure*}
\begin{equation*}
  \begin{array}{rclr}
    F & - & \multicolumn{2}{l}{\mbox{a role logic formula}} \mnl
    A & - & \multicolumn{2}{l}{\mbox{unary predicate}} \mnl
    f & - & \multicolumn{2}{l}{\mbox{binary predicate}} \mnl
    \procedure & ::= & \procName ( \unaryList ) = \statement \mnl
    \refinement & ::= & \procName \implies \procName \mnl
    \unaryList & ::= & A \mid \unaryList, A \mnl
    \statement & ::= & \assignmentStat
    & \cmnt{assignment statement} \mnls
    & \mid & \procName ( \paramList )
    & \cmnt{procedure call} \mnls
    & \mid & \assume\, F
    & \cmnt{assume statement} \mnls
    & \mid & \assert\, F
    & \cmnt{assert statement} \mnls
    & \mid & \spec\, F_E
    & \cmnt{specification} \mnls
    & \mid & \statement \lor \statement 
    & \cmnt{non-deterministic choice} \mnls
    & \mid & \statement \land \statement 
    & \cmnt{conjunction} \mnls
    & \mid & \statement ; \statement 
    & \cmnt{sequential composition} \mnl    
    \assignmentStat & ::= & A \becomes F
    & \cmnt{update of unary predicate} \mnls
    & \mid & F_1 . f \becomes F_2 
    & \cmnt{update of binary predicate} \mnls
    & \mid & F_1 . \twid{f} \becomes F_2
    & \cmnt{update of inverse of binary predicate} \mnl
    F_E & ::= & \multicolumn{2}{l}{A \mid f \mid \ID \mid F_1 \land F_2 \mid \lnot F} \mnls
        & \mid & \multicolumn{2}{l}{F' \mid \twid{F} \mid \card{{\geq}k} F} \mnls
        & \mid & \assignmentStat \mid \modify\, \items \mid \procName(\paramList) \mnl
    \paramList & ::= & F \mid \paramList, F \mnl
    \items & ::= & \modItem \mid \items, \modItem \mnl
    \modItem & ::= & A \subecomes F
    & \cmnt{modification of unary predicate} \mnls
    & \mid & F_1 . f \subecomes F_2
    & \cmnt{modification of binary predicate} \mnls
    & \mid & F_1 . \twid{f} \subecomes F_2
    & \cmnt{modification of inverse of binary predicate}
  \end{array}
\end{equation*}
\caption{Syntax of a Small Imperative 
  Language\label{fig:imperativeLanguage}}
\end{figure*}

\begin{figure}
  \begin{equation*}
    \begin{array}{rclr}
      \tr{P_1 \implies P_2} & = &
      \begin{array}[t]{@{}l}
        (\tr{S_1} \land \\
        \ \lnot{\tr{S_2}(B_1 \mapsto A_1,\ldots, B_n \mapsto A_n)}) \\
        \mbox{ is not satisfiable, where:} \mnl
        \begin{array}[t]{l}
          P_1(A_1,\ldots,A_n) = S_1 \mnl
          P_2(B_1,\ldots,B_n) = S_2 \mnl
          \tr{S_2} \mbox{ has no fresh predicates} \mnl
        \end{array} \\
      \end{array} \\
      \tr{A \becomes F} & = &
      \squareb{A \iff \oldv{F}} \land \modUnary\, A \mnl
      \tr{F_1.f \becomes F_2} & = &
      \begin{array}[t]{@{}l}
        \squareb{\oldv{F_1} \implies \squareb{f \iff \oldv{F_2}}} \ \land \\
        \squareb{\lnot \oldv{F_1} \implies \squareb{f \iff \oldv{f}}}\ \land \\
        \modBinary\ f \mnl
      \end{array} \\
      \tr{F_1.\twid{f} \becomes F_2} & = &
      \begin{array}[t]{@{}l}
        \squareb{\oldv{F_1} \implies \squareb{\twid{f} \iff \oldv{F_2}}} \ \land \\
        \squareb{\lnot \oldv{F_1} \implies \squareb{\twid{f} \iff \twid{\oldv{f}}}}\ \land \\
        \modBinary\ f \mnl
      \end{array} \\
      \tr{P(F_1,\ldots,F_n)} & = & 
      \begin{array}[t]{@{}l}
         \tr{S}(A_1 \mapsto \oldv{F_1}, \ldots, A_n \mapsto \oldv{F_n}) \mnl
         \mbox{where } P(A_1,\ldots,A_n) = S \mnl
       \end{array} \\
      \tr{\assume\, F} & = & \oldv{F} \land \skipStat \mnl
      \tr{\assert\, F} & = & \oldv{F} \implies \skipStat \mnl
      \tr{\spec\, F} & = & \tr{F} \mnl
      \tr{s_1 \land s_2} & = & \tr{s_1} \land \tr{s_2} \mnl
      \tr{s_1 \lor s_2} & = & \tr{s_1} \lor \tr{s_2} \mnl
      \tr{s_1 ; s_2} & = & 
      \begin{array}[t]{@{}l}
        \formRen{\epsilon}{k}{\tr{s_1}}\ \land \mnl
        (\squareb{\lnot \errorPred} \implies \formRen{0}{k}{\tr{s_2}}) \mnl
        k - \mbox{fresh element of $\{ 1, 2, \ldots \}$} \mnl
      \end{array} \\
      \tr{\modify\ E} & = & \mtr{E} \mnl
      \modUnary\, A & \equiv & 
      \begin{array}[t]{@{}l}
        \bigwedge_{B \neq A} \squareb{B \iff \oldv{B}} \ \land \mnl
        \bigwedge_{g} \squareb{\squareb{g \iff \oldv{g}}} \ \land \mnl
        \squareb{\errorPred \iff \oldv{\errorPred}} \mnl
      \end{array} \\
      \modBinary f & \equiv & 
      \begin{array}[t]{@{}l}
        \bigwedge_{B} \squareb{B \iff \oldv{B}} \ \land \mnl
        \bigwedge_{g \neq f} \squareb{\squareb{g \iff \oldv{g}}}\ \land \mnl
        \squareb{\errorPred \iff \oldv{\errorPred}} \mnl
      \end{array} \\
      \skipStat & \equiv & 
      \begin{array}[t]{@{}l}
        \bigwedge_{B} \squareb{B \iff \oldv{B}} \ \land \mnl
        \bigwedge_{g} \squareb{\squareb{g \iff \oldv{g}}}\ \land \mnl
        \squareb{\errorPred \iff \oldv{\errorPred}} \mnl
      \end{array}
    \end{array}
  \end{equation*}  
  \caption{Predicates Describing the Semantics
  of the Language from Figure~\label{fig:statementPredicates}}
\end{figure}

\smartparagraph{Program state.}  The state of the program is
a first-order structure interpreting the language $L = \cala
\cup \calf$ where $\cala$ is a finite set of unary
predicates and $\calf$ is a finite set of binary predicates.
We fix a countable universe of objects $\obj$, and assume
that each structure has the same universe $\obj$.  To
specify the structure, it suffices to give the set $e A
\subseteq \obj$ for each unary predicate $A \in \cala$, and
a binary relation $e f \subseteq \obj \times \obj$ for each
binary predicate $f \in \calf$.

\smartparagraph{Extended language.}  For each $k \in \{
\epsilon, 0, 1, \ldots \}$ we define the language $\ifresh{L}{k}$.  We
identify $\ifresh{L}{\epsilon}$ with $L$,
$\ifresh{A}{\epsilon}$ with $A$ and $\ifresh{f}{\epsilon}$
with $f$.  For $k \in \{ 0,1, \ldots \}$, we let
$\ifresh{A}{k}$ be a fresh unary predicate symbol, and
$\ifresh{f}{k}$ a fresh binary predicate symbol, and
$\ifresh{L}{k}$ be the set of all $\ifresh{A}{k}$ and
$\ifresh{f}{k}$.  The notation $\formRen{i}{j}{F}$ for $i, j
\in \{ \epsilon, 0, 1, 2\ldots \}$ denotes a formula
resulting from $F$ by replacing all elements of $\ifresh{L}{i}$ with the
corresponding elements of $\ifresh{L}{j}$.

\smartparagraph{Describing relations in the extended
  language.}  The meaning of each statement in our
imperative language is a binary relation on $L$-structures.
We describe a binary relation on structures with an $\RLtwo$
formula in the language $\ifresh{L}{0} \cup
\ifresh{L}{\epsilon}$.  The predicates in
$\ifresh{L}{\epsilon}$ denote the state components in the
final state; the predicates in $\ifresh{L}{0}$ denote the
state components in the initial state.  If $F$ is a formula
in language $\ifresh{L}{\epsilon}$, then $\oldv{F}$ is a
shorthand for the formula $\formRen{\epsilon}{0}{F}$ in the
language $\ifresh{L}{0}$; the purpose of $\oldv{F}$ is to
denote the value of the formula $F$ evaluated in the initial
state.

Define the renaming operator $\strucRen{i}{j}{}$ such that
if $\ifresh{e}{i}$ is an $\ifresh{L}{i}$-structure, then
$\ifresh{e}{j} = \strucRen{i}{j}{\ifresh{e}{i}}$ is an
$\ifresh{L}{j}$-structure such that $\ifresh{e}{j}\,
\ifresh{A}{j} = \ifresh{e}{i}\, \ifresh{A}{i}$ and
$\ifresh{e}{j}\, \ifresh{f}{j} = \ifresh{e}{i}\,
\ifresh{f}{i}$ for all $A, f \in L$.  Then the relation on
$L$-structures denoted by an $\RLtwo$ formula $F$ in
language $\ifresh{L}{0} \cup \ifresh{L}{\epsilon}$ is $\{
\tu{e,e'} \mid \tr{F}((\strucRen{\epsilon}{0}{e})\ \cup\ e')
\}$.

\smartparagraph{Assignment statements.}  The imperative
language in Figure~\ref{fig:statementPredicates} contains
three forms of assignment statements.

The statement $A \becomes F$ evaluates to the formula $F$,
which denotes a unary predicate.  The statement makes $A$
true precisely for those object for which $F$ was true in
the initial state.  Unary predicates other than $A$ as well
as binary predicates remain unchanged.

The statement $F_1.f \becomes F_2$ generalizes the statement
$x.f=y$ in a language like Java by allowing simultaneous
modification of fields of a set of objects.  Formula $F_1$
specifies the set of objects whose fields are modified.
Formula $F_2$ specifies the new value of the field $f$ for
objects in $F_1$.  Unary predicates and binary predicates
other than $f$ remain unchanged.  Note that $F_2$ may
specify a relation, which is particularly interesting when
$F_1$ denotes a set with more then one element because it
allows the value of the field to depend on the source object
of the field.  As a special case, $F_1.f \becomes g$ copies
the entire field $g$ into field $f$ for all objects in the
set given by $F_1$, and, in particular, $\boolTrue.f
\becomes g$ copies the field $g$ into $f$.  The statement
$F_1. \twid{f} \becomes F_2$ is dual to $F_1.f \becomes F_2$, 
and updates the inverse of the predicate $f$.

\smartparagraph{Statements for specification.}  The
statement $\assume\, F$ filters out the state transitions
for which $F$ does not hold in the initial state.  The
statement $\assert\, F$ behaves arbitrarily if the condition
given by $F$ does not hold in the initial state.  The state
contains an additional predicate $\errorPred$, which makes
it easier to detect that an arbitrary behavior occurred (the
sequential composition operator ensures that the
$\errorPred$ value is propagated).  

The statement $\spec\, F_E$ allows describing relations on
states directly in terms of an extended $\RLtwo$ formula
$F_E$.  Formula $F_E$ allows assignment statements and
modifies statements in addition to the constructs of
$\RLtwo$.  The relation symbols of $\RLtwo$ may refer to
relation symbols of the extended language, which allows
stating relations between pre and postcondition.  We also
allow non-recursive procedure calls in the specification
when they expand to constructs not containing sequential
composition.

\smartparagraph{$\modify$ specifications.} The construct
\begin{equation*}
  \modify\ e_1,\ldots,e_n
\end{equation*}
is useful for specifying frame conditions.  Each expression
$e_i$ specifies a set of possible modifications.  Any finite
number of modifications can occur as the result of the
action specified by the $\modify$ specification.

\begin{figure*}
  \begin{equation*}
    \begin{array}{l}
      \mtr{\modify\, e_1,\ldots,e_n} = \mnl
      \begin{array}[t]{l}
        \mbox{\bf let } \{ e_1,\ldots,e_n \} = \mnl
        \{ \begin{array}[t]{@{}l}
          A_1 \subecomes F_1, \ldots, A_k \subecomes F_k, \mnl
          F_{k+1}.f_{k+1} \subecomes G_{k+1}, \ldots, F_l.f_l \subecomes G_l, \mnl
          F_{l+1}.\twid{f_{l+1}} \subecomes G_{l+1}, \ldots, F_m.\twid{f_m} \subecomes G_m \} \mnl
        \end{array} \\
        \mbox{\bf in} \mnl
        \bigwedge\limits_{A \notin \{ A_1,\ldots,A_k \}} 
        \squareb{A \miff \oldv{A}} \ \land \mnl
        \bigwedge\limits_{A \in \{ A_1,\ldots,A_k \}}
        \squareb{(\lnot \oldv{A} \land 
        \bigwedge_{A_i \equiv A} \lnot \oldv{F_i}) 
        \implies \lnot A} \ \land \mnl
        \bigwedge\limits_{f \notin \{f_{k+1},\ldots,f_m\}}
        \squareb{\squareb{f \miff \oldv{f}}} \ \land \mnl
        \bigwedge\limits_{f \in \{f_{k+1},\ldots,f_m\}}
        [[(
          \bigwedge\limits_{\begin{array}{@{}c}
              \scriptstyle
              f_i \equiv f \\
              \scriptstyle
              i \leq l 
            \end{array}} \lnot F_i' \ \land\
          \bigwedge\limits_{\begin{array}{@{}c}
              \scriptstyle
              f_i \equiv f \\
              \scriptstyle
              l < i
            \end{array}} \lnot F_i)\ \implies \
          (f \miff \oldv{f})]] \mnl
        \bigwedge\limits_{f \in \{f_{k+1},\ldots,f_m\}}
        [[(
          \lnot \oldv{f} \land 
          \bigwedge\limits_{\begin{array}{@{}c}
              \scriptstyle
              f_i \equiv f \\
              \scriptstyle
              i \leq l 
            \end{array}} \lnot (F_i' \land G_i) \land
          \bigwedge\limits_{\begin{array}{@{}c}
              \scriptstyle
              f_i \equiv f \\
              \scriptstyle
              l < i
            \end{array}} \lnot (F_i \land \twid{G_i}))\ \implies \
          \lnot f]]
      \end{array}
    \end{array}
  \end{equation*}
  \caption{Semantics of $\modify$ statement.\label{fig:modify}}
\end{figure*}

\begin{figure}
\begin{verbatim}
proc assignClients() = 
spec old(GlobalInvariant) =>
 (modify WaitingClients, AssignedClients, 
         old(WaitingClients).server :<= Servers,
         Servers.clients :<= old(WaitingClients)) &
 !{WaitingClients} &
 [AssignedClients <=> 
           old(AssignedClients | WaitingClients)] &
 GlobalInvariant

proc assignOneClient(cl) =
spec old(GlobalInvariant) & 
     [cl => old(WaitingClients)] =>
 (modify WaitingClients, AssignedClients, 
         cl.server :<= Servers,
         Servers.clients :<= cl) &
 [WaitingClients | cl <=> old(WaitingClients)] &
 [AssignedClients <=> old(AssignedClients) | cl] &
 GlobalInvariant
\end{verbatim}
  \caption{Specifications for $\assignClients$ and $\assignOneClient$
  extended with side effect specifications.
  \label{fig:modifyExample}}
\end{figure}

\begin{example}
  Figure~\ref{fig:modifyExample} shows the specifications
  $\assignClients$ and $\assignOneClient$ from
  Figure~\ref{fig:exaProgram} extended with frame-condition
  specifications.  The frame condition for
  $\assignOneClient$ specifies that only the sets
  $\WaitingClients$ and $\AssignedClients$ can change, which
  is useful if the system contains some additional set of
  objects, such as a set $\ProcessedClients$.  Next, the
  frame-condition specifies that the only binary relations
  that were modified are $\server$ and $\clients$.  The
  modifies expression $(\Servers.\clients \subecomes
  \clVar)$ indicates that the the only way in which the
  $\clients$ relation is changed is by introducing an edge
  from a $\Servers$ object to the $\clVar$ object, or by
  removing an edge from a $\Servers$ object.  (The removal
  of the edge does not, in fact, occur in
  $\assignOneClientIMPL$ in Figure~\ref{fig:exaProgram}, but
  the frame condition is a conservative approximation.)  The
  amount of detail in specifications such as modifies
  clauses depends on how strong property we need to prove.
  The strength of the property, in turn, depends either on
  some high-level program correctness requirement, or on the
  amount of information we need about the procedure to prove
  the properties of its callers.  In
  Figure~\ref{fig:exaProgram}, we did not use $\modify$
  specification for $\assignOneClient$ because we did not
  need it to prove the conformance of $\assignClientsIMPL$
  with respect to $\assignClients$.  However, even in
  Figure~\ref{fig:exaProgram} we needed to know that, for
  example, $\getServer$ preserves the global invariant,
  which follows from the fact that it does not modify any
  sets or relations (the conjunction with $\skipStat$
  implies that $\getServer$ is a pure function).
\end{example}

In general, there are three forms of modification
expressions.  The expression $A \subecomes F$ specifies
modifications that remove an element from the set $A$ or
insert into $A$ an element that satisfies $F$.  For example,
after executing the statement
\begin{equation*}
  \modify\, A \subecomes F
\end{equation*}
the set $A$ may contain any subset of the set of objects
given by the expression $\oldv{A} \lor F$.  The expression
$F_1.f \subecomes F_2$ specifies modifications that 1)
remove a tuple $\tu{o_1,o_2}$ from the relation interpreting
the predicate $f$, when $o_1$ satisfies $F_1$, or 2) insert
a tuple $\tu{o_1,o_2}$ into the relation interpreting $f$,
when $o_1$ satisfies $F_1$ and $\tu{o_1,o_2}$ satisfies $F_2$.
Similarly, $F_1.\twid{f} \subecomes F_2$ allows removing
$\tu{o_1,o_2}$ from the interpretation of $f$ when $o_1$
satisfies $F_1$, or inserting $\tu{o_1,o_2}$ when $o_2$
satisfies $F_1$ and $\tu{o_1,o_2}$ satisfy $\twid{F_2}$.

If $r_i$ is the relation describing a modification
given by the expression $e_i$, then the meaning
of $\modify\, e_1,\ldots,e_n$ is given by the relation
\begin{equation} \label{eqn:simpleSemantics}
  (r_1 \cup \ldots \cup r_n)^{*}
\end{equation}
where $r^{*}$ denotes the transitive closure of relation
$r$.  The simple semantics~(\ref{eqn:simpleSemantics})
provides good intuition about the meaning of $\modify$
statement and makes it clear that the $\modify$ statement is
idempotent
\cite{KuncakLeino03InPlaceRefinementEffectChecking}.
Figure~\ref{fig:modify} presents an alternative semantics,
which directly encodes a modify statement as an $\RLtwo$
formula.  The advantage of the semantics in
Figure~\ref{fig:modify} is that it eliminates the need for
transitive closure of the transition relation.



\smartparagraph{Disjunction and conjunction.}  The language allows
computing disjunction and conjunction on statements.
Disjunction $\lor$ has a natural interpretation as a
non-deterministic choice of commands.  Conjunction $\land$
is useful for combining nondeterministic statements.
Logical operations on statements translate directly to the
corresponding logical operations on $\RLtwo$ formulas.

\smartparagraph{Computing sequential composition.}  When
encoding sequential composition of statements in $\RLtwo$,
we introduce copies $\ifresh{L}{i}$ of predicate names in
$L$ for $i \in \{ 1, 2, \ldots \}$.  These copies of
predicate names denote the values of predicates at program
points between the initial and the final program state.
Because the definition of relation composition $r_1 \circ
r_2 = \{ \tu{x,z} \mid \exists y.\ \tu{x,y} \in r_1 \land
\tu{y,z} \in r_2 \}$ involves existential quantification
over $y$, we treat the newly introduced predicates as being
existentially quantified.  The technique of introducing new
predicate names allows us to precisely compute relation
composition even for non-deterministic commands.

\smartparagraph{Procedure calls.}  The meaning of a
procedure is also a relation on states, where the initial
state is extended with one unary predicate symbol for each
parameter name.  In the simple translation of
Figure~\ref{fig:statementPredicates}, a procedure call
identifies parameters with the sets that describe their
values by performing the substitution.  Substitution
suffices to give semantics to procedures because we assume
that the recursion is split using refinement claims.  Loops
are represented as recursive procedures, so we effectively
require loop invariants.

\smartparagraph{Refinement claims.}  If $P_1$ and $P_2$ are
procedure names, the refinement claim $P_1 \implies P_2$ is
a proof obligation that the relation given by the body of
procedure $P_1$ is contained in the relation given by the
body of $P_2$.  The intended use of the refinement claim is
the specification procedure summaries, which allows breaking
the cycles in the call graphs of mutually recursive
procedures.  Figure~\ref{fig:statementPredicates} shows how
each refinement claim reduces to a test whether an $\RLtwo$
formula is satisfiable.  When generating the $\RLtwo$
formula, we rename the parameters of $P_2$ replacing them
with the corresponding parameters of $P_1$.

To ensure that the satisfiability test treats newly
introduced predicates as existentially quantified, we impose
a restriction that the translation $\tr{S_2}$ contains no
newly introduced predicates from $\ifresh{L}{i}$ for $i \in
\{ 1, 2, \ldots \}$.  We impose this restriction because
$\tr{S_2}$ appears under negation in the satisfiability
test, so newly introduced predicates in $\tr{S_2}$ would be
universally quantified, thus violating the semantics of
sequential composition for non-deterministic statements.
The restriction on $S_2$ is satisfied when $S_2$ contains no
sequential composition, which is typically the case for a
large class of procedure summaries.  

By providing sufficiently many procedure summaries, the
partial correctness of a program is reduced to a finite
number of refinement claims.  By discharging these claims
using a decision procedure for $\RLtwo$, we decide the
partial correctness of the program.

\smartparagraph{Fixpoint computation.}  If some procedure
summaries are not supplied by the programmer, they can be
inferred using fixpoint computation.  An algorithm for
fixpoint computation can be derived from the fixpoint
semantics of mutually recursive procedures using abstract
interpretation \cite{CousotCousot77AbstractInterpretation,
  CousotCousot79SystematicDesignProgramAnalysis,
  CousotCousot77RecursiveProcedures,
  YiHarrison93AutomaticGenerationInterproceduralAnalyses}.
A special case of this approach is to select a finite subset
of all $\RLtwo$ formulas and define a lattice structure on
the set using the entailment of formulas.  A simple way to
define a finite subset of formulas is to consider only
$\RLtwo$ formulas with quantifier depth at most $k$, for
some $k \geq 1$.  Boolean shape analysis constraints in
Section~\ref{sec:bsac} have quantifier depth at most two, so
they can be used as a basis of fixpoint computation.



  \subsection{Describing Boolean Shape Analysis Constraints}
\label{sec:bsac}

\begin{figure}
  \begin{equation*}
    \begin{array}{rcl}
      F & ::= & \curlyb{C} \mid \curlyb{\curlyb{C_1' \land C_2 \land R}} 
         \mid F_1 \land F_2 \mid \lnot F \mnl
      C & ::= & A \mid C_1 \land C_2 \mid \lnot C \mnl
      R & ::= & f \mid \lnot f \mid R_1 \lor R_2 \mnl
      A & - & \mbox{atomic unary predicate} \mnl
      f & - & \mbox{atomic binary predicate}
    \end{array}
  \end{equation*}
  \caption{Boolean Shape Analysis Constraints expressed as a sublogic of 
    $\RLtwo$\label{fig:bsacSyntax}}
\end{figure}

Boolean Shape Analysis Constraints
\cite{KuncakRinard03OnBooleanAlgebraShapeAnalysisConstraints}
are a natural language for describing dataflow facts of
shape analyses \cite{SagivETAL02Parametric}.

Figure~\ref{fig:bsacSyntax} presents the syntax of Boolean
Shape Analysis Constraints as a subset of role logic.  This
presentation of Boolean Shape Analysis Constraints shows
that they are a subset of the decidable fragment $\RLtwo$ of
role logic.  In fact, Boolean Shape Analysis Constraints do
not use counting quantifiers, so they are already
expressible in the two-variable predicate logic $L^2$
(without counting).



\paragraph{A note on usability of role logic.}  An anecdotal
evidence of the usability of role logic is the fact that all
results of
\cite{KuncakRinard03OnBooleanAlgebraShapeAnalysisConstraints}
were initially shown using role logic notation and then
translated into the standard first-order logic notation.  We
have found the variable-free aspect of role logic convenient
when showing the results of
\cite{KuncakRinard03OnBooleanAlgebraShapeAnalysisConstraints}.
We have subsequently discovered the connection of role
logic with $C^2$
\cite{GraedelETAL97TwoVariableLogicCountingDecidable}, presented
in Section~\ref{sec:decidable}, and the connection with
description logics \cite{BaaderETAL03DescriptionLogicHandbook}, presented
in Section~\ref{sec:dlencoding}.


  \subsection{Encoding an Expressive Description Logic}
\label{sec:dlencoding}

\begin{figure}
  \begin{equation*}
    \begin{array}{rcl}
      C & ::= & A \mid C \sqcap C \mid \lnot C \mid
                {\geq} n R. C \mnl
      R & ::= & f \mid R \sqcap R \mid \lnot R \mid U \mid
                R^{-1} \mid \restr{R}{C} \mid \idof{C} \mnl
      A & - & \mbox{atomic unary predicate} \mnl
      f & - & \mbox{atomic binary predicate}
    \end{array}
  \end{equation*}
  \caption{An Expressive Description Logic\label{fig:dl}}
\end{figure}

\begin{figure}
  \begin{equation*}
    \begin{array}{rcl}
      \tr{A} & = & A \mnl
      \tr{C_1 \sqcap C_2} & = & \tr{C_1} \land \tr{C_2} \mnl
      \tr{\lnot C} & = & \lnot \tr{C} \mnl
      \tr{{\geq} n R.C} & = & \card{{\geq}n} (\tr{R} \land \tr{C}) \\
      \\
      \tr{f} & = & f \mnl
      \tr{R_1 \sqcap R_2} & = & \tr{R_1} \land \tr{R_2} \mnl
      \tr{\lnot R} & = & \lnot \tr{R} \mnl
      \tr{U} & = & \boolTrue \mnl
      \tr{R^{-1}} & = & \twid{\tr{R}} \mnl
      \tr{\restr{R}{C}} & = & \tr{R} \land \tr{C} \mnl
      \tr{\idof{C}} & = & \ID \land \tr{C}
    \end{array}
  \end{equation*}
  \caption{Translation of an Expressive Description Logic to Role Logic with
    Two Variables\label{fig:dlToRL}}
\end{figure}

Figure~\ref{fig:dl} presents an Expressive Description Logic
fragment where roles have no transitive operators
\cite[Chapter 5]{BaaderETAL03DescriptionLogicHandbook}.
Figure~\ref{fig:dlToRL} presents the translation of the
Expressive Description Logic into $\RLtwo$.  The translation
maps the concepts $C$ and roles $R$ of description logic
into unary and binary predicates of role logic.  The
translation to $\RLtwo$ in Figure~\ref{fig:dlToRL} implies
that the description logic in Figure~\ref{fig:dl} is
decidable.  The fact that interesting description logics can
be translated to $\RLtwo$ is not surprising once we have
established that $\RLtwo$ and $C^2$ have equal expressive
power.  Nevertheless, it is interesting to observe the
simplicity of the translation from the description logic to
$\RLtwo$, which is partly because both description logic and
role logic avoid explicit occurrences of variables.  

Using rules
\begin{equation*}
  \begin{array}{rcl}
    \tr{R_1 \circ R_2} & = & \tr{R_1} \circ \tr{R_2} \mnl
    \tr{R^{*}} & = & \tr{R}^{*}
  \end{array}
\end{equation*}
we can translate operations on binary relations into the
full role logic, but not into the decidable fragment
$\RLtwo$.  Decidability of interesting description logics
that contain transitive closure but do not have tree model
property is an open problem \cite[Page
214]{BaaderETAL03DescriptionLogicHandbook}.

\paragraph{A note on terminology.} The term ``role'' 
has different meanings in different formalisms for
describing structures.  In \cite{KuncakETAL02RoleAnalysis},
a role corresponds to a unary predicate (set), in
description logics
\cite{BaaderETAL03DescriptionLogicHandbook}, a role
corresponds to a binary predicate (relation), and in
entity-relationship diagrams in databases
\cite{Chen76EntityRelationshipModel}, a role corresponds to
a position $i$ ($1 \leq i \leq n$) in a $n$-tuples of an
$n$-ary relation.  To avoid the confusion, we use the
well-established terms of $n$-ary ``predicate'' (or
``relation''), keep the name ``role logic'' for the logic
described in Figure~\ref{fig:RoleLogicSyntax}, because the
term ``role logic'' appears appropriate regardless of the
particular interpretation of the word ``role''.

\newcommand{\dlfrag}{{\cal DL}{-}\{ \mbox{\bf trans}, \mbox{\bf compose} \}}
\paragraph{Description Logics Corresponding to $C^2$.}\footnote{Note added on 31 October 2003, after
  becoming aware of
  \cite{Borgida96RelativeExpressivenessDescriptionLogicsPredicateLogics}.}
The result \cite[Theorem
4]{Borgida96RelativeExpressivenessDescriptionLogicsPredicateLogics}
reports that the description logic without transitive
closure and relation composition (denoted $\dlfrag$)
corresponds precisely to $C^2$.  The results of
Section~\ref{sec:decidable} and
\cite{Borgida96RelativeExpressivenessDescriptionLogicsPredicateLogics}
imply that our logic $\RLtwo$ has the same expressive power
as $\dlfrag$.  One of the differences between $\RLtwo$ and
$\dlfrag$ is that $\RLtwo$ contains the prime operator $F'$
and does not contain the $\mbox{\bf product}$ operation of
$\dlfrag$.  Another difference is the foundation of role
logic on de Bruijn lambda calculus notation, as described in
Section~\ref{sec:recipe}.



\section{Related Work}
\label{sec:related}

We have initially developed role logic to provide a
foundation for role analysis \cite{KuncakETAL02RoleAnalysis,
  Kuncak01DesigningRoleAnalysis}.  We have subsequently
studied a simplification of role analysis constraints and
showed a characterization of such constraints using formulas
\cite{KuncakRinard02TypestateCheckingRegularGraphConstraints}.
Parametric analysis based on three-valued logic was
introduced in \cite{SagivETAL99Parametric,
  SagivETAL02Parametric} with interprocedural analysis in
\cite{RinetzkySagiv01InterprocedualShapeAnalysis} and
application to abstract data type verification in
\cite{LevAmiETAL00PuttingStaticAnalysisWorkVerification}.  A
characterization of dataflow facts used for shape analysis
was presented in
\cite{Yorsh03LogicalCharacterizationsHeapAbstractions,
  KuncakRinard03OnBooleanAlgebraShapeAnalysisConstraints}.
A decidable logic for expressing connectivity properties of
the heap was presented in
\cite{BenediktETAL99LogicForLinked}.

Specifying the semantics of programs using predicates dates
back to axiomatic program semantics
\cite{Hoare69AxiomaticBasisComputerProgramming,
  Floyd67AssigningMeaningsPrograms}.  An approach that
uses a first-order logic theorem prover tailed for program verification
is \cite{FlanaganETAL02ExtendedStaticCheckingJava}.

Like \cite{KlarlundSchwartzbach94Transductions,
  KlarlundSchwartzbach93GraphTypes,
  JensenETAL97MonadicLogic, Moeller01PALE}, in
Section~\ref{sec:transductions} we use an expressive yet
decidable logic to encode fragments of straight-line code.
Our approach differs primarily in using logic $\RLtwo$ over
general graphs whose decidability follows from the
decidability of $C^2$, where
\cite{KlarlundSchwartzbach94Transductions,
  KlarlundSchwartzbach93GraphTypes,
  JensenETAL97MonadicLogic, Moeller01PALE} uses graph types
whose decidability follows from the decidability of monadic
second-order logic over trees.  We expect that these two
logics can be combined in a fruitful way.

We have extended our language with constructs that make it
possible to directly express higher-level state
transformations, which is the idea related to the chemical
reaction model of \cite{FradetMetayer97ShapeTypes,
  FradetMetayer98StructuredGamma}, the verification of
database transactions
\cite{BenediktETAL98VerifiablePropertiesDatabaseTransactions},
the simultaneous assignments of \cite{Moeller01PALE}, and in
wide-spectrum languages
\cite{Morgan94ProgrammingFromSpecifications,
  BackWright98RefinementCalculus}.  Verification of a form
of modifies clauses using a theorem prover was presented
\cite{LeinoETAL02DataGroupsSideEffects,
  KuncakLeino03InPlaceRefinementEffectChecking}.  Further
approaches to pointer and shape analysis include
\cite{ChongRugina03StaticAnalysisAccessedRegionsRecursiveDataStructures,
  WhaleyRinard99CompositionalPointerEscapeAnalysis,
  ChaseETAL90AnalysisPointersStructures,
  GhiyaHendren96TreeOrDag,
  FradetETALStaticVerificationPointer,
  GaugneETAL96StaticDetectionPointerErrors,
  WilsonLam95PointerAnalysis}.

Description logics
\cite{BaaderETAL03DescriptionLogicHandbook,
  Borgida95DescriptionLogicsDataManagement} share many of
the properties of role logic and have been traditionally
applied to knowledge bases.  It is likely that description
logics can be used for shape analysis as well.  It would be
particularly interesting to consider description logics with
transitive operators, whose decidability is related to the
decidability of dynamic logic
\cite{HarelETAL00DynamicLogic}.  
Reasoning about
the satisfiability of expressive description logics over all
structures and over finite structures is presented in
\cite{Calvanese96FiniteModelReasoningDescriptionLogics,
  Calvanese96UnrestrictedFiniteModelReasoning}.  Reasoning
about entity-relationship diagrams
\cite{Chen76EntityRelationshipModel} is presented in
\cite{LenzeriniNobili87SatisfiabilityDependencyConstraintsER}.
Some connections between object models and heap invariants
are presented in
\cite{KuncakRinard01ObjectModelsHeapsInterpretations,
  Jackson01OMInvariants}.

Like the Alloy modelling language
\cite{Jackson02AlloyTOSEM}, role logic combines the notation
of predicate calculus with the notation of relational
algebras.  It may be possible to combine the notation of
Alloy with the notation of role logic, and to combine the
benefits of bounded model checking used in Alloy Analyzer
with the benefits of a decision procedure for $\RLtwo$.

A recent approach to reasoning about mutable imperative data
structure is separation logic
\cite{IshtiaqOHearn01BIAssertionLanguage,
  Reynolds00IntuitionisticReasoningMutable,
  Reynolds02SeparationLogic,
  CalcagnoETAL00SemanticAnalysisPointerAliasings,
  CalcagnoETAL02DecidingValiditySpatialLogicTrees}.  We are
currently working on integrating some aspects of spatial
logic to support more flexible notation for records in role
logic.

Interactive theorem provers have also been used for
reasoning about dynamically allocated data structures
\cite{MehtaNipkow03ProvingPointerProgramsHigherOrderLogic,
  BackETAL03ReasoningPointersRefinementCalculus}; it may be
interesting to incorporate a decision procedure for $\RLtwo$
into these general tools.




\section{Conclusions}
\label{sec:conclusions}

We believe that role logic notation is a convenient way of
expressing properties of first-order structures.
First-order structures are a natural way to model the state
in object-oriented programs, or a the state of a knowledge
base or a database.  Role logic can be combined with
traditional variable-based notation in a natural way.
Furthermore, interesting subsets of role logic are
decidable.  Decision procedures for role logic can therefore
enable shape analysis of programs and have similar benefits
as description logics in knowledge bases.


\paragraph{Acknowledgements}
We thank Patrick Lam for useful discussions, comments on the
paper, and an implementation of an early version of role
logic normalization algorithm in Fall 2001, we thank Andreas
Podelski for discussion of using formulas to perform shape
analysis, we thank Thomas Reps for discussions on
summarizing procedures using two-vacabulary structures, we
thank C.  Scott Ananian for discussion of a draft of this
paper in Spring 2003, we thank Derek Rayside, Mooly Sagiv,
and Greta Yorsh for useful discussions, and Darko Marinov
for comments on the paper.



\bibliographystyle{plain}
\bibliography{pnew}

\end{document}


